\documentclass[oneside,11pt]{article}

\usepackage[utf8]{inputenc}
\usepackage[T1]{fontenc}

\usepackage{amsthm}
\usepackage{amsmath}
\usepackage{amsfonts}
\usepackage{amssymb}

\usepackage{graphicx} 
\usepackage{graphbox}
\usepackage{float}
\usepackage{booktabs}
\usepackage{multirow}
\usepackage{rotating}
\usepackage{array}
\usepackage{xcolor}
\usepackage{subcaption}
\usepackage[ruled]{algorithm2e}

\newcolumntype{C}[1]{>{\centering\arraybackslash}p{#1}}
\newcolumntype{R}[1]{>{\raggedleft\let\newline\\\arraybackslash\hspace{0pt}}m{#1}}
\newcolumntype{L}[1]{>{\raggedright\let\newline\\\arraybackslash\hspace{0pt}}m{#1}}

\usepackage{breakcites}

\usepackage{nowidow}

\usepackage{enumitem}

\usepackage[top=1.5cm,bottom=2cm,right=2.5cm,left=2.5cm]{geometry}
\usepackage{titling}

\usepackage{natbib}

\usepackage{ifplatform}

\usepackage{textcomp}
\usepackage{bbm}

\ifwindows
\fi

\allowdisplaybreaks

\newcommand{\rd}{\mathrm{d}}

\newcommand{\bm}{\mathbf{m}}

\newcommand{\defin}{:=}

\DeclareFontFamily{OT1}{pzc}{}
\DeclareFontShape{OT1}{pzc}{m}{it}{<-> s * [1.10] pzcmi7t}{}
\DeclareMathAlphabet{\mathpzc}{OT1}{pzc}{m}{it}

\newtheorem{theorem}{Theorem}[section]

\newtheorem{proposition}{Proposition}[section]

\newtheorem{example}{Example}[section]

\renewenvironment{proof}[1]{\textit{Proof#1.}}{\qed\\} 

\usepackage[pdftex,final,bookmarksnumbered,bookmarksopen=false,breaklinks,colorlinks]{hyperref}
\hypersetup{
	final,
	bookmarksnumbered=true,
	bookmarksopen=true,
	bookmarksopenlevel=0,
	unicode=false,
	pdftoolbar=true,
	pdfmenubar=true,
	pdffitwindow=false,
	pdftitle={A family of toroidal diffusions with exact likelihood inference},
	pdfdisplaydoctitle=true,
	pdftoolbar=true,
	pdfmenubar=true,
	pdflang={English},
	pdfauthor={Eduardo Garcia-Portugues, Michael Sørensen},
	pdfsubject={arXiv paper},
	pdfcreator={Eduardo Garcia-Portugues},
	pdfproducer={Eduardo Garcia-Portugues},
	pdfkeywords={Angular data}{Diffusion bridges}{Stochastic processes}{Torus},
	pdfnewwindow=true,
	breaklinks=true,
	hidelinks,
	linkcolor=black,
	citecolor=black,
	filecolor=magenta,
	urlcolor=cyan
}

\newif\ifmain
\maintrue
\newif\ifsupplement
\supplementtrue

\newif\iffigstabs
\figstabstrue

\usepackage{bm}
\newcommand{\bmx}{\bm{x}}
\newcommand{\bmy}{\bm{y}}
\newcommand{\bmz}{\bm{z}}
\newcommand{\bma}{\bm{a}}
\newcommand{\bms}{\bm{s}}
\newcommand{\bmk}{\bm{k}}
\newcommand{\bmn}{\bm{n}}
\newcommand{\bmm}{\bm{m}}
\newcommand{\bmmu}{\bm{\mu}}
\newcommand{\bmtheta}{\bm{\theta}}

\newcommand{\bmSigma}{\bm{\Sigma}}
\newcommand{\bmxi}{\bm{\xi}}
\newcommand{\bmbeta}{\bm{\beta}}
\newcommand{\bzero}{\bm{0}}
\newcommand{\bmX}{\bm{X}}
\newcommand{\bmW}{\bm{W}}
\newcommand{\bmV}{\bm{V}}
\newcommand{\bmU}{\bm{U}}
\newcommand{\bmTheta}{\bm{\Theta}}
\newcommand{\bmI}{\bm{I}}
\newcommand{\bmB}{\bm{B}}
\newcommand{\bmD}{\bm{D}}
\newcommand{\bmR}{\bm{R}}
\newcommand{\bmM}{\bm{M}}
\newcommand{\bmN}{\bm{N}}

\newcommand{\bmK}{\bm{K}}
\newcommand{\bmcalI}{\bm{\mathcal{I}}}
\newcommand{\bmcalJ}{\bm{\mathcal{J}}}
\newcommand{\bmcalK}{\bm{\mathcal{K}}}
\def\sect/{Section }
\newcommand{\abbr}[1]{}
\def\sde/{stochastic differential equation}
\def\sdes/{stochastic differential equations}
\def\tpd/{transition probability density}
\def\tpds/{transition probability densities}
\def\pdf/{density}
\def\pdfs/{densities}
\def\cdf/{distribution function}
\def\kdes/{kernel density estimators}
\def\dgp/{data generation process}
\def\dgps/{data generation processes}

\begin{document}

\ifmain

\title{A family of toroidal diffusions with exact likelihood inference}
\setlength{\droptitle}{-1cm}
\predate{}%
\postdate{}%
\date{}

\author{Eduardo Garc\'ia-Portugu\'es$^{1,3}$ and Michael Sørensen$^{2}$}
\footnotetext[1]{Department of Statistics, Carlos III University of Madrid (Spain).}
\footnotetext[2]{Department of Mathematical Sciences, University of Copenhagen (Denmark).}
\footnotetext[3]{Corresponding author. e-mail: \href{mailto:edgarcia@est-econ.uc3m.es}{edgarcia@est-econ.uc3m.es}.}
\maketitle

\begin{abstract}
We provide a class of diffusion processes for continuous time-varying multivariate angular data with explicit transition probability densities, enabling exact likelihood inference. The presented diffusions are time-reversible and can be constructed for any pre-specified stationary distribution on the torus, including highly-multimodal mixtures. We give results on asymptotic likelihood theory allowing one-sample inference and tests of linear hypotheses for $k$ groups of diffusions, including homogeneity. We show that exact and direct diffusion bridge simulation is possible too. A class of circular jump processes with similar properties is also proposed. Several numerical experiments illustrate the methodology for the circular and two-dimensional torus cases. The new family of diffusions is applied (\textit{i}) to test several homogeneity hypotheses on the movement of ants and (\textit{ii}) to simulate bridges between the three-dimensional backbones of two related proteins.
\end{abstract}
\begin{flushleft}
\end{flushleft}

\section{Introduction}

The modelling of time-varying angular observations on the circle $\mathbb{T}^1\defin[0,2\pi)$, with $0$ and $2\pi$ identified, has been mainly focused on processes in discrete time \citep[see][\sect/10]{Pewsey2021}, with some seminal contributions being the Markov process of \cite{Wehrly1980} and the autoregressive processes of \cite{Breckling1989} and \cite{Fisher1994}; see \cite{Harvey2024} for a recent score-driven approach to the latter.

One of the first continuous-time processes on $\mathbb{T}^1$ is the \emph{von Mises process} by \cite{Kent1975}. The model describes the evolution of an angle $\{\Theta_t\}\subset\mathbb{T}^1$ by the stochastic differential equation
\begin{align}
	\rd \Theta_t=\frac{\sigma^2\kappa}{2} \sin(\mu-\Theta_t)\rd t+\sigma \rd W_t,\label{eq:vMKent}
\end{align}
where $\{W_t\}$ is a Wiener process, $\kappa\geq0$ controls the strength of the drift towards $\mu\in[0,2\pi)$, and $\sigma>0$ is the diffusion coefficient. This Ornstein--Uhlenbeck-like process is ergodic with the von Mises distribution being its stationary distribution. The von Mises process has been applied in mathematical biology and physics, see, e.g., \cite{Hill1997}, \cite{Codling2005}, and \citet[\sect/5.3.3]{Frank2005}. Unlike the Ornstein--Uhlenbeck process, no closed-form expression for the transition probability density of \eqref{eq:vMKent} is known, which hampers likelihood-based inference.

\cite{Garcia-Portugues2019} generalized the diffusion \eqref{eq:vMKent} to \emph{Langevin diffusions} on the $p$-dimensional torus $\mathbb{T}^p=[0,2\pi)^p$, $p\geq1$. These processes are time-reversible, ergodic, and can be defined by any differentiable probability density function on $\mathbb{T}^p$ (in particular, a wrapped normal density) that acts as the stationary density. However, the transition probability density is not explicitly known for the studied Langevin diffusions, with likelihood-based estimation requiring either expensive numerical solutions of the Kolmogorov's forward equation, faster but less accurate pseudo-likelihoods, or more precise likelihood approximations that are specific to the wrapped normal diffusion. \cite{Golden2017} used the bivariate wrapped normal diffusion for their hidden Markov model for protein structure evolution because of its faster and more accurate estimation within the available time-reversible and ergodic diffusions on the torus.

In the present paper, we make several contributions towards advancing diffusion models on the circle and torus, particularly targeting the lack of exact likelihood inference. Our main contribution is a family of diffusion processes on the torus for which closed-form expressions for the transition probability density functions can be given, implying that the likelihood function is explicitly known. The new family is constructed through a transformation argument specific to the geometry of $\mathbb{T}^p$, which we present first in \sect/\ref{sec:circ} for the (technically easier) case $p=1$ and then in \sect/\ref{sec:torus} for $p\geq1$. The diffusions in the new family are time-reversible and ergodic, and can have any differentiable density on the torus as the stationary density while maintaining the closed-form nature of the transition density. We derive asymptotic likelihood theory for samples observed at discrete times, both for the one sample case and for linear hypotheses in the case of $k$ groups, in \sect/\ref{sec:likinf}. The asymptotic results are validated with several numerical experiments. The new family of diffusions allows exact bridge simulation too, for which we give a direct algorithm in \sect/\ref{sec:bridge}. We take a new direction in \sect/\ref{sec:jump} to adapt the construction technique to produce jump processes on the circle with a given stationary distribution, analytic likelihood function, and exact bridge simulation. Finally, we give two in-depth biological applications of the new diffusion inferential toolbox in \sect/\ref{sec:data}. The first uses homogeneity tests to investigate the effect of several factors on the circular ant movement induced by wall-following behaviour. The second exploits diffusion bridges to generate data-driven bridges between the protein backbones of a calcium-free calmodulin and a Ca$^{2+}$-calmodulin, using the dihedral-pairs representation of proteins.

\section{Circular case}
\label{sec:circ}

We consider the time-homogeneous diffusion given by the stochastic differential equation \abbr{ (sde)}
\begin{align}
	\rd X_t = -\frac{\sigma^2f'(X_t)}{2f(X_t)^3}\rd t+\frac{\sigma}{f(X_t)}\rd W_t,\label{eq:diff}
\end{align}
where $\{W_t\}$ is a Wiener process, $\sigma>0$ is a volatility coefficient, and $f:\mathbb{R}\to\mathbb{R}^+$ is a differentiable \emph{circular} probability density function\abbr{ (pdf)}, that is, a function satisfying (\textit{i}) $f(x+2k\pi)=f(x)$ for any $x\in\mathbb{R}$ and $k\in\mathbb{Z}$ and (\textit{ii}) \smash{$\int_{0}^{2\pi} f(\theta)\,\rd\theta = 1$}. The wrapping of $\{X_t\}$ by the $2\pi$-modulus operator produces the diffusion $\Theta_t = X_t \mod 2\pi$, which (by definition) solves the \sde/
\begin{align}
	\rd \Theta_t = -\frac{\sigma^2f'(\Theta_t)}{2f(\Theta_t)^3}\rd t+\frac{\sigma}{f(\Theta_t)}\rd W_t\label{eq:diffcirc}
\end{align}
on the circle $\mathbb{T}^1=[0,2\pi)$. The diffusion \eqref{eq:diff} is not ergodic. Its speed measure is infinite, so no stationary \pdf/ exists. However, the wrapped process \eqref{eq:diffcirc} is ergodic. The solution to \eqref{eq:diffcirc} with $\Theta_0 = \theta_0 \in \mathbb{T}^1$ is given by
\begin{align}
	\Theta_t=F^{-1}(\sigma W_t+F(\theta_0))\mod 2\pi,\label{eq:sol}
\end{align}
where $F:\mathbb{R}\to\mathbb{R}$, $F(x)\defin\int_{0}^x f(s)\,\rd s$, is a circular cumulative distribution function. Particularly, $F$ is a one-to-one function verifying $F(x)= F(x\mod 2\pi) + \lfloor x/(2\pi)\rfloor$ and having inverse $F^{-1}(y) = F^{-1}(y\mod 1) + 2\pi\lfloor y\rfloor$, for all $x,y\in\mathbb{R}$. That $F$ is a \emph{circular} \cdf/ makes a remarkable difference with respect to the regular (linear) case. Indeed, if $g$ is a regular \pdf/ on $\mathbb{R}$ with associated \cdf/ $G:\mathbb{R}\to[0,1]$, then $G^{-1}(\sigma W_t+G(x_0))$ is an \emph{explosive} solution for the diffusion~\eqref{eq:diff} in which $f$ is replaced by $g$, as $\sigma W_t+G(x_0)$ will eventually exit $[0,1]$.

The following result collects the main properties of $\{\Theta_t\}$, in particular the transition probability density\abbr{ (tpd)}, i.e., the \pdf/ of $\Theta_{t_2}\mid \Theta_{t_1}=\theta_1$. Proofs of all results can be found in the Appendix.

\begin{proposition}\label{prp:Xt}
	Define $\Theta_t \defin F^{-1}(\sigma W_t+F(\theta_0))\mod 2\pi$, $t>0$. Then:
	\begin{enumerate}
		\item $\{\Theta_t\}$ solves \eqref{eq:diffcirc} with $\Theta_0 = \theta_0 \in \mathbb{T}^1$.
		\item For $t_2>t_1>0$ and $\theta_1\in\mathbb{T}^1$, the \tpd/ of $\{\Theta_t\}$ is
		\begin{align}
			p_{t_2-t_1}(\theta_2\mid \theta_1)
			=2\pi f_{\mathrm{WN}}(2\pi F(\theta_2);2\pi F(\theta_1),4\pi^2(t_2-t_1)\sigma^2)f(\theta_2), \; \theta_2\in\mathbb{T}^1,\label{eq:tpd}
		\end{align}
		where $f_{\mathrm{WN}}(\theta;\mu,\sigma^2)\defin\sum_{k\in\mathbb{Z}}\phi_{\sigma^2}(\theta-\mu+2k\pi)$ is the wrapped normal \pdf/, with $\phi_{\sigma^2}(\cdot)$ being the \pdf/ of $\mathcal{N}(0,\sigma^2)$.
		\item $\{\Theta_t\}$ is ergodic with stationary \pdf/ $f$ and is time-reversible.
	\end{enumerate}
\end{proposition}

Some comments on the connections of diffusion \eqref{eq:diffcirc} with other processes follow. First, it is easy to check that \eqref{eq:diffcirc} is a Langevin diffusion on the circle \citep[\sect/2]{Garcia-Portugues2019} with variable diffusion coefficient given by $\sigma/f$ and stationary \pdf/ $f$. This variable diffusion coefficient also relates \eqref{eq:diffcirc} to \cite{Roberts2002}'s Langevin \emph{tempered} diffusions used to simulate from densities with untractable normalizing constants. Indeed, in \eqref{eq:diffcirc} $f$ can be replaced with an unnormalized \pdf/ $f^*\propto f$ and still produce an ergodic diffusion with stationary \pdf/ $f$, as the normalizing constant is absorbed by $\sigma$. There is also an interesting relation of the diffusion \eqref{eq:diffcirc} to \cite{Wehrly1980}'s (discrete-time) Markov process on the circle, as \eqref{eq:tpd} has a similar structure to the transition probability density in \cite{Wehrly1980}'s Equation (3). Additionally, the transformation \eqref{eq:sol} leading to \eqref{eq:diffcirc} resembles that in \cite{Forman2014}, which is based on transforming a process $X_t$ with stationary \cdf/ $\Pi$ into $\tau(X_t)$ with $\tau\defin G^{-1}\circ\Pi$ and $G$ a target stationary \cdf/. However, the construction \eqref{eq:sol} cannot be cast in terms of $\tau$ as $\{W_t\}$ is not ergodic and, if replacing it with an ergodic process in $[0,1]$, the transformation would yield a diffusion on $[0,2\pi)$ with dynamics that do not allow it to move freely around the circle, as a barrier would be created at $0\equiv2\pi$.

Some notable examples of \eqref{eq:diffcirc} follow.

\begin{example}[Circular Brownian motion] \label{ex:0}
	For the circular uniform density $f\equiv 1/(2\pi)$, \eqref{eq:diffcirc} becomes $\rd \Theta_t = \tilde{\sigma}\rd W_t$, the Brownian motion on the circle with volatility $\tilde{\sigma}\defin 2\pi\sigma$. This volatility gives a useful reference to compare the scale of $\sigma$ for a non-uniform \pdf/ $f$.
\end{example}

\begin{example}[A von Mises diffusion] \label{ex:1}
	The von Mises distribution $\mathrm{vM}(\mu,\kappa)$ has \pdf/ $f_{\mathrm{vM}}(\theta;\mu,\kappa)%
	\defin(2\pi I_0(\kappa))^{-1}\allowbreak
	\exp\{\kappa \cos(\theta-\mu)\}$, where $I_0$ is the modified Bessel function of the first kind and zeroth order. Thus \eqref{eq:diffcirc} has the form
	\begin{align}
		\rd \Theta_t%
		=&\; \frac{\sigma^2\kappa \sin(\Theta_t-\mu)}{2\exp\{2\kappa \cos(\Theta_t-\mu)\}}\rd t+\frac{\sigma}{\exp\{\kappa \cos(\Theta_t-\mu)\}}\rd W_t\label{eq:diffvm}
	\end{align}
	if the normalizing constant is absorbed in~$\sigma$. The diffusion \eqref{eq:diffvm} shares with \eqref{eq:vMKent} the same stationary \pdf/ $\mathrm{vM}(\mu,\kappa)$, yet the latter does not have a known \tpd/.Besides, in \eqref{eq:vMKent} the stationary distribution is caused by the drift only, while in \eqref{eq:diffvm} it is the combined effect of the drift and the diffusion coefficient. %
\end{example}

\begin{example}[A wrapped Cauchy diffusion] \label{ex:2}
	The Wrapped Cauchy distribution $\mathrm{WC}(\mu,\rho)$ has \pdf/ $f_{\mathrm{WC}}(\theta;\mu, \rho)%
	\defin(2 \pi)^{-1} (1-\rho^{2})
	(1+\rho^{2}-2 \rho \cos (\theta-\mu))^{-1}$. This \pdf/ follows by transforming with a $2\pi$-modulus the Cauchy distribution $\mathrm{C}(\mu,\sigma^2)$ with \pdf/ $f_{\mathrm{C}}(x;\mu,\sigma^2)\defin\sigma\pi^{-1}((x-\mu)^2+\sigma^2)^{-1}$. Specifically,
	$f_{\mathrm{WC}}(\theta;\mu, \rho)=\sum_{k\in\mathbb{Z}} f_{\mathrm{C}}(\theta+2k\pi;\mu,\sigma^2)$, with $\rho\defin e^{-\sigma}$. Absorbing the normalizing constant in $\sigma$, \eqref{eq:diffcirc} produces a diffusion with stationary distribution $\mathrm{WC}(\mu,\rho)$:
	\begin{align*}
		\rd \Theta_t%
		=&\; \sigma^2\rho\sin (\Theta_t-\mu)(1+\rho^{2}-2 \rho \cos (\Theta_t-\mu))^2\rd t+\sigma(1+\rho^{2}-2 \rho \cos (\Theta_t-\mu))\rd W_t.
	\end{align*}
\end{example}

Figure \ref{fig:1} shows the multimodal \tpds/ of the diffusions in Examples \ref{ex:1}--\ref{ex:2}, as well as for a diffusion given by a mixture of von Mises densities. The \tpds/ stabilize about the stationary \pdf/.

\begin{figure}[!ht]
	\centering
	\hspace*{-0.25cm}\includegraphics[height=0.235\textheight]{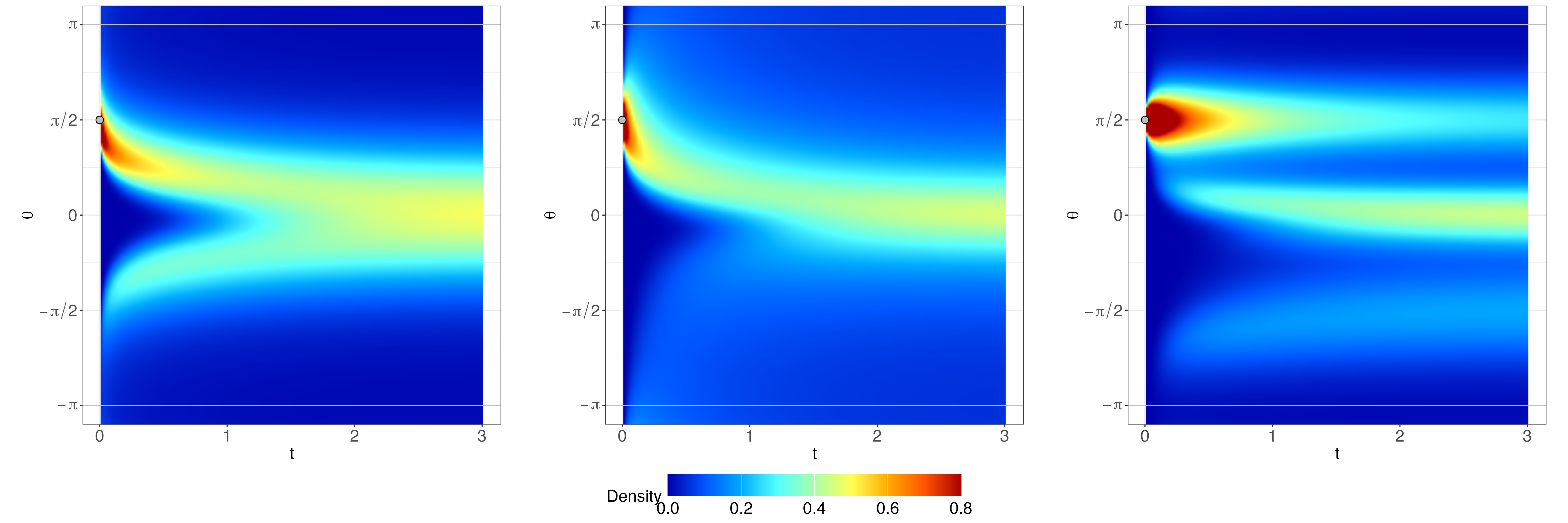}
	\caption{\small Temporal evolution of the \tpd/ $p_t(\cdot\mid \theta_0)$ for three circular diffusions \eqref{eq:diffcirc}. From left to right, the diffusions are given by the distributions $\mathrm{vM}(0,2)$, $\mathrm{WC}(0,0.5)$, and the mixture $0.4\mathrm{vM}(0,8)+0.3\mathrm{vM}(-\pi/2,3)+0.3\mathrm{vM}(\pi/2,5)$. In all the plots, $\theta_0=\pi/2$, $T=3$, and $\sigma=0.25$.}
	\label{fig:1}
\end{figure}

\section{Toroidal case}
\label{sec:torus}

The circular diffusion \eqref{eq:diffcirc} can be extended to the $p$-dimensional torus, $\mathbb{T}^p=[0,2\pi)^p$, $p\geq1$, with $0$ and $2\pi$ again identified. In analogy with the case $p=1$, we start with a differentiable toroidal \pdf/ $f$, i.e.\ a differentiable function $f:\mathbb{R}^p\to\mathbb{R}^+$ such that (\textit{i}) $f(\bmx+2\bmk\pi)=f(\bmx)$ for all $\bmx\in\mathbb{R}^p$ and $\bmk\in\mathbb{Z}^p$ and (\textit{ii}) \smash{$\int_{\mathbb{T}^p} f(\bmtheta)\,\rd\bmtheta = 1$}. The joint toroidal \cdf/ $F(\bmx)\defin\int_{\bzero}^{\bmx} f(\bms)\,\rd \bms$ does not provide a one-to-one function of $\mathbb{R}^p$ onto itself. Instead we use \cite{Rosenblatt1952}'s transformation to yield a one-to-one map $R:\mathbb{R}^p\to\mathbb{R}^p$. Define the conditional densities $f_j(x_j \mid x_1, \ldots, x_{j-1}) = f(x_1, \ldots, x_j)/f(x_1, \ldots, x_{j-1})$ (with a slight abuse of notation for the marginal distribution). Clearly, $f_j(x_j + 2 k_j \pi \mid x_1 + 2 k_1 \pi, \ldots, x_{j-1} +2 k_{j-1} \pi) = f_j(x_j \mid x_1, \ldots, x_{j-1})$ for all $\bmx\in\mathbb{R}^j$ and $\bmk\in\mathbb{Z}^j$, $j=2,\ldots,p$. Now, a one-to-one map $R:\mathbb{R}^p\to\mathbb{R}^p$ is given by
\begin{align*}
	R(\bmx)%
	\defin(F_1(x_1),F_2(x_2\mid x_1),\ldots,F_p(x_p\mid x_1,\ldots,x_{p-1}))^\prime,%
\end{align*}
where $F_1$ is the first marginal \cdf/ of $F$, and $F_j(x_j \mid x_1, \ldots , x_{j-1})=\int_{0}^{x_j} f(s \mid x_1, \ldots, x_{j-1})\,\rd s$, $j=2,\ldots,p$.
Let $R^{-1}$ denote the inverse function, which for $p=2$ has the form $R^{-1}(\bmy) = (F^{-1}_1(y_1), F_2^{-1}(y_2\mid F^{-1}_1(y_1)))^\prime$.
As defined, $R$ is a one-to-one map satisfying $R(\bmx)=R(\bmx\mod2\pi)+\lfloor\bmx/(2\pi)\rfloor$ and $R^{-1}(\bmy)=R^{-1}(\bmy\mod 1)+2\pi\lfloor\bmy\rfloor$, for all $\bmx,\bmy\in\mathbb{R}^p$, where the modulus and floor functions are applied componentwise.

We define the extension of the circular diffusion \eqref{eq:diffcirc} to $\mathbb{T}^p$ as
\begin{align}
	\rd \bmTheta_{t}=\frac{1}{2}\begin{pmatrix}
		\mathrm{tr}[\bmSigma(\mathrm{H}R_1^{-1})(R(\bmTheta_{t}))]\\
		\vdots\\
		\mathrm{tr}[\bmSigma(\mathrm{H}R_p^{-1})(R(\bmTheta_{t}))]
	\end{pmatrix}\rd t+(\mathrm{D}R^{-1})(R(\bmTheta_{t}))\bmSigma^{1/2}\rd \bmW_{t},\label{eq:difftorus}
\end{align}
which is obtained by wrapping the diffusion $\{\bmX_t\}$ analogously defined on $\mathbb{R}^p$. Above, $\mathrm{D}$ and $\mathrm{H}$ stand for the Jacobian matrix and Hessian matrix operators, $\bmSigma$ is a covariance matrix, and $\{\bmW_t\}$ is a standard $p$-variate Wiener process. Note that the first $q$ coordinates of the drift vector depend only on the first $q$ coordinates of $\bmTheta_t$, and that the matrix $(\mathrm{D}R^{-1})(R(\bmTheta_{t}))$ is a lower triangular matrix. When $p=2$ and $\bmSigma=\sigma^2\bmI_2$, with $\bmI_p$ denoting the $p\times p$ identity matrix, \eqref{eq:difftorus} simplifies to
\begin{align}
	\begin{pmatrix}
		\rd \Theta_{1,t}\\
		\rd \Theta_{2,t}
	\end{pmatrix}=&-\frac{\sigma^2}{2}%
	\begin{pmatrix}
		f_1'(\Theta_{1,t})(f_1(\Theta_{1,t}))^{-3} \\
		f_2'(\Theta_{2,t}\mid \Theta_{1,t}) (f_2(\Theta_{2,t}\mid \Theta_{1,t}))^{-3}+b(\bmTheta_t)
	\end{pmatrix}\rd t\nonumber\\
	&+ \sigma \begin{pmatrix}
		(f_1(\Theta_{1,t}))^{-1} & 0\\
		-h(\bmTheta_{t}) & (f_2(\Theta_{2,t}\mid \Theta_{1,t}))^{-1}
	\end{pmatrix}\begin{pmatrix}
		\rd W_{1,t}\\
		\rd W_{2,t}
	\end{pmatrix},\label{eq:difftorus2}
\end{align}
where $b(\bmTheta_t)=(\partial_{x_1}h(\bmTheta_{t})) (f_1(\Theta_{1,t}))^{-1} - (\partial_{x_2}h(\bmTheta_{t}))h(\bmTheta_{t})$ and
\begin{align*}
	h(x_1,x_2)\defin\frac{\partial{x_1} F_2(x_2\mid x_1)}{f(x_1,x_2)} = \frac{1}{f_1(x_1)f(x_1,x_2)}\left[\partial^2_{x_1}F(x_1,x_2) - \frac{f_1'(x_1)}{f_1(x_1)}\partial_{x_1}F(x_1,x_2)\right].
\end{align*}
The expression above follows by tedious but straightforward computations involving the implicit function theorem. A simple instance of \eqref{eq:difftorus} for $p\geq2$ is when $F(\bmx)=F_1(x_1)\cdots F_p(x_p)$, which gives the drift-independent and volatility-dependent diffusion
\begin{align}
	\rd \bmTheta_{t}=\frac{1}{2}\begin{pmatrix}
		-\sigma_1^2 f_1'(\Theta_{1,t})(f_1(\Theta_{1,t}))^{-3}\\
		\vdots\\
		-\sigma_p^2 f_p'(\Theta_{p,t})(f_p(\Theta_{p,t}))^{-3}
	\end{pmatrix}\rd t+\mathrm{diag}(f_1(\Theta_{1,t}),\ldots,f_p(\Theta_{p,t})))^{-1} \bmSigma^{1/2}\rd \bmW_{t} \label{eq:diffindep}
\end{align}
for an unrestricted covariance matrix $\bmSigma$.

The following result collects the main properties of $\{\bmTheta_t\}$.

\begin{proposition}\label{prp:Xt2}
	Define $\bmTheta_t \defin R^{-1}(\bmSigma^{1/2} \bmW_t + R(\bmtheta_0)) \mod 2\pi$, $t>0$. Then:
	\begin{enumerate}
		\item $\{\bmTheta_t\}$ solves \eqref{eq:difftorus} with $\Theta_0 = \theta_0 \in \mathbb{T}^p$.
		\item For $t_2 > t_1 > 0$ and $\bmtheta_1\in\mathbb{T}^p$, the \tpd/ of $\{\bmTheta_t\}$ is
		\begin{align}
			p_{t_2-t_1}(\bmtheta_2\mid \bmtheta_1)
			=(2\pi)^p f_{\mathrm{WN}}(2\pi R(\bmtheta_2);2\pi R(\bmtheta_1),4\pi^2(t_2-t_1)\bmSigma)f(\bmtheta_2), \; \bmtheta_2\in\mathbb{T}^p,\label{eq:btpd}
		\end{align}
		where $f_{\mathrm{WN}}(\bmtheta;\bmmu,\bmSigma)\defin\sum_{\bmk\in\mathbb{Z}^p}\phi_{\bmSigma}(\bmtheta-\bmmu+2\bmk\pi)$ is the multivariate wrapped normal \pdf/, with $\phi_{\bmSigma}(\cdot)$ being the \pdf/ of $\mathcal{N}_p(\bzero,\bmSigma)$.
		\item $\{\bmTheta_t\}$ is ergodic with stationary \pdf/ $f$ and is time-reversible.
	\end{enumerate}
\end{proposition}

The examples below extend Examples \ref{ex:0}--\ref{ex:1}. An analogous extension of Example \ref{ex:2} could be possible using the bivariate wrapped Cauchy model by \cite{Kato2015a}, but such an extension is, to the best of the authors' knowledge, (currently) hampered by the absence of readily usable formulae for its conditional densities.

\begin{example}[Toroidal Brownian motion]
	For the toroidal uniform density $f\equiv 1/(2\pi)^p$, \eqref{eq:diffindep} yields the Brownian motion on $\mathbb{T}^p$ with covariance matrix $\tilde{\bmSigma}\defin (2\pi)^{2p}\bmSigma$, since $\rd \bmTheta_t = \tilde{\bmSigma}^{1/2}\rd \bmW_t$.
\end{example}

\begin{example}[A bivariate von Mises diffusion] \label{ex:3}
	A random vector $(\Theta_1,\Theta_2)^\prime$ on $\mathbb{T}^2$ has the bivariate (sine) von Mises distribution \citep{Singh2002}, $\mathrm{BvM}(\mu_1,\mu_2,\kappa_1,\kappa_2,\lambda)$, if its \pdf/~is
	\begin{align}
		f_{\mathrm{BvM}}(\theta_1,\theta_2;\mu_1,\mu_2,\kappa_1,\kappa_2,\lambda)\defin&\;C(\kappa_1,\kappa_2,\lambda)\exp\{\kappa_1\cos(\theta_1-\mu_1)+\kappa_2\cos(\theta_2-\mu_2)\nonumber\\
		&\qquad\qquad\qquad\qquad+\lambda\sin(\theta_1-\mu_1)\sin(\theta_2-\mu_2)\},\label{eq:diffBVM}
	\end{align}
	where the normalizing constant is \smash{$C(\kappa_1,\kappa_2,\lambda)^{-1}=\allowbreak 4 \pi^{2} \sum_{m=0}^{\infty}\binom{2m}{m} (\lambda^{2}/(4\kappa_{1} \kappa_{2}))^m I_{m}(\kappa_{1}) I_{m}(\kappa_{2})$} and \smash{$(\mu_1,\mu_2)^\prime\in\mathbb{T}^2$, $\kappa_1,\kappa_2\geq 0$}, and $\lambda\in\mathbb{R}$. Define $\mu_{2\lambda}(\theta_1)\defin\mu_2+\tan^{-1}((\lambda/\kappa_2)\sin(\theta_1-\mu_1))$ and $\kappa_{2\lambda}(\theta_1)\defin[\kappa_2^2+\lambda^2\sin(\theta_1-\mu_1)^2]^{1/2}$. The marginal and conditional \pdfs/ of $\Theta_1$ and $\Theta_2\mid \Theta_1=\theta_1$ are, respectively,
	\begin{align*}
		f_1(\theta_1)&= C(\kappa_1,\kappa_2,\lambda)^{-1} 2\pi I_0(\kappa_{2\lambda}(\theta_1))\exp\{\kappa_1\cos(\theta_1-\mu_1)\},\\
		f_2(\theta_2\mid \theta_1)&=f_{\mathrm{vM}}(\theta_2;\mu_{2\lambda}(\theta_1),\kappa_{2\lambda}(\theta_1)),
	\end{align*}
	and can be readily plugged into \eqref{eq:difftorus2}.
\end{example}

Mixtures of bivariate von Mises distributions can also be used to define the diffusion \eqref{eq:difftorus2}. Figure \ref{fig:2} displays snapshots of the \tpd/ of such diffusion for different times, showing the convergence to the stationary mixture \pdf/.

\begin{figure}[ht!]
	\centering
	\hspace*{-0.25cm}\includegraphics[height=0.235\textheight]{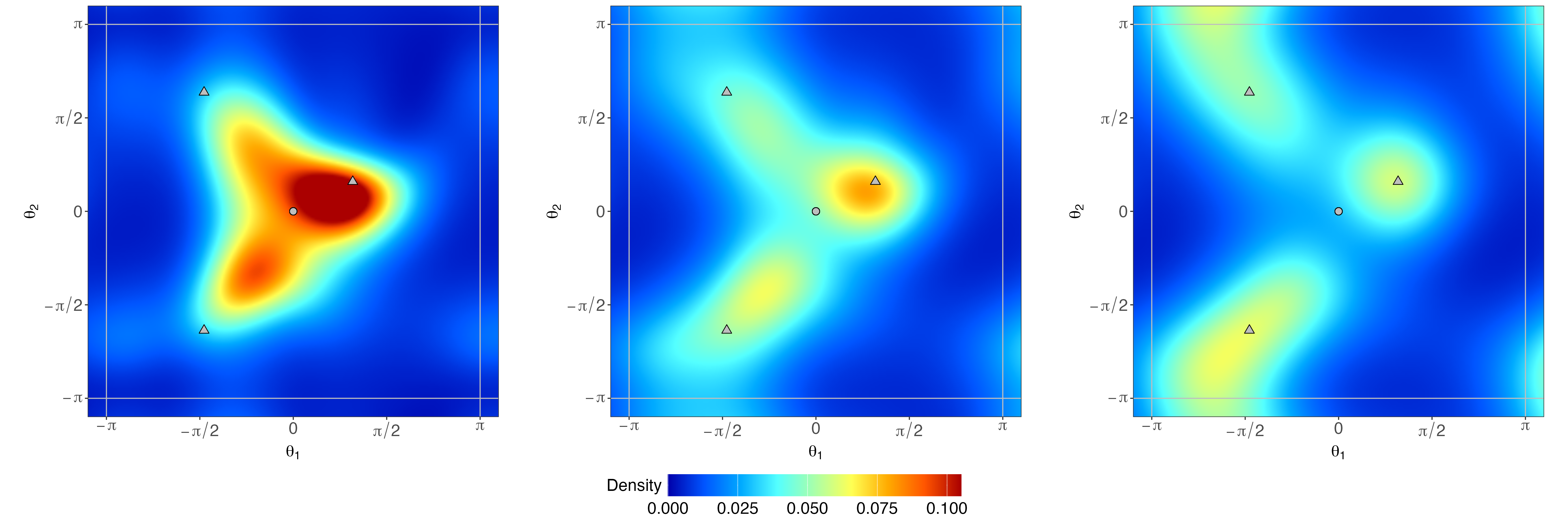}
	\caption{\small Temporal snapshots of the \tpd/ $p_t(\cdot\mid \bmtheta_0)$ for $t=0.2,0.4,1$ (from left to right) on the toroidal diffusion \eqref{eq:difftorus2} given by the mixture $0.4\mathrm{BvM}(-1.5,2,1,1,-0.5)+0.4\mathrm{BvM}(-1.5,-2,1,1.5,1)+0.2\mathrm{BvM}(1,0.5,2,2,0)$. In all the plots, $\bmtheta_0=(0,0)^\prime$ (grey circle) and $\sigma=0.5$. The grey triangles represent the mean of the mixture components.}
	\label{fig:2}
\end{figure}

\section{Likelihood inference}
\label{sec:likinf}

\subsection{One-sample inference}
\label{sec:one}

Suppose we have a parametrized class of circular \pdfs/ $f_{\bmbeta}$, $\bmbeta \in B \subseteq \mathbb{R}^{q-1}$, and that we want to draw inference about the $q$-dimensional parameter $\bmxi^\prime = (\bmbeta^\prime,\sigma) \in \Xi = B \times \mathbb{R}^+$ on the basis of the discrete-time observations $\Theta_0, \Theta_\Delta, \ldots, \Theta_{n\Delta}$, $\Delta>0$, of the diffusion given by \eqref{eq:diffcirc}. To simplify the exposition, we consider only the circular case. Similar results for the general toroidal case can be obtained in the same way.

By the Markov property, the log-likelihood function is
\begin{align}
	\ell_n(\bmxi; \{\Theta_{i\Delta}\}_{i=0}^n) \equiv \ell_n(\bmxi) = \sum_{i=1}^n \log p_\Delta (\Theta_{i\Delta};\bmxi \mid  \Theta_{(i-1)\Delta}) \label{eq:ll}
\end{align}
where
\begin{align*}
	p_\Delta (\theta_2;\bmxi \mid  \theta_1) = f_{\bmbeta}(\theta_2) \sum_{k\in\mathbb{Z}} \phi_{\sigma^2\Delta} (h_k(\theta_1,\theta_2;\bmbeta))
\end{align*}
with $h_k(\theta_1,\theta_2;\bmbeta) = F_{\bmbeta}(\theta_2) - F_{\bmbeta}(\theta_1) + k$. We denote $L_n(\bmxi)\defin\exp(\ell_n)$ to the likelihood of $\bmxi$.

Under the conditions of Theorem \ref{prp:mle} below, the likelihood function is twice continuously differentiable with respect to $\bmxi$. The score function is the vector-valued function
\begin{align}
	\label{eq:score}
	\partial_{\bmxi} \ell_n(\bmxi) = \sum_{i=1}^n \partial_{\bmxi} \log p_\Delta (\Theta_{i\Delta};\bmxi\mid  \Theta_{(i-1)\Delta})
\end{align}
where, for $i=1,\ldots,q-1$,
\begin{align*}
	\partial_{\xi_i} \log p_\Delta (\theta_2;\bmxi \mid \theta_1) &= \partial_{\beta_i} \log f_{\bmbeta}(\theta_2) %
	\\ & \hspace{-2cm} -\frac{\partial_{\beta_i}F_{\bmbeta}(\theta_2) - \partial_{\beta_i}F_{\bmbeta}(\theta_1)}{\sigma^2\Delta}\frac{\sum_{k\in\mathbb{Z}} h_k(\theta_1,\theta_2;\bmbeta) \phi_{\sigma^2\Delta} (h_k(\theta_1,\theta_2;\bmbeta))}{\sum_{k\in\mathbb{Z}} \phi_{\sigma^2\Delta} (h_k(\theta_1,\theta_2;\bmbeta))}.
\end{align*}
The last parameter coordinate $\xi_q$ is $\sigma$ and
\begin{align*}
	\partial_\sigma \log p_\Delta (\theta_2;\bmxi\mid\theta_1) = -\frac{1}{\sigma} + \frac{1}{\sigma^3 \Delta }\frac{\sum_{k\in\mathbb{Z}} h_k(\theta_1,\theta_2;\bmbeta)^2 \phi_{\sigma^2\Delta} (h_k(\theta_1,\theta_2;\bmbeta))}{\sum_{k\in\mathbb{Z}} \phi_{\sigma^2\Delta} (h_k(\theta_1,\theta_2;\bmbeta))}.
\end{align*}

\begin{theorem}\label{prp:mle}
	Assume the following conditions hold:
	\begin{enumerate}[label=\arabic{*}.,ref=\arabic{*}]
		\item %
		$\bmbeta \mapsto f_{\bmbeta}(\theta)$ and $\bmbeta \mapsto F_{\bmbeta}(\theta)$ are twice continuously differentiable for all $\theta \in \mathbb{T}^1$.\label{prp:mle:1}
		\item $f_{\bmbeta}>0$, and $\partial_{\bmbeta}^{k} f_{\bmbeta}$ and $\partial^{k}_{\bmbeta}F_{\bmbeta}$
		are continuous functions of $(\bmbeta,\theta)$ for $k=0,1,2$.\label{prp:mle:2}

		\item The interior of $\Xi$ is non-empty and contains $\bmxi_0$, the true value of the parameter.\label{prp:mle:3}

		\item The covariance matrix of $\partial_{\bmxi} \log p_\Delta (\Theta_{2\Delta};\bmxi\mid \Theta_\Delta)$ under the stationary distribution,
		\begin{align}
			\label{eq:Fisher}
			\bmcalI^\Delta(\bmxi) = \mathrm{E}_{\bmxi}(\partial_{\bmxi} \log p_\Delta (\Theta_{2\Delta};\bmxi\mid \Theta_\Delta)\partial_{\bmxi} \log p_\Delta (\Theta_{2\Delta};\bmxi\mid \Theta_\Delta)^\prime),
		\end{align}
		is positive definite for all $\bmxi \in \Xi$.  ($\mathrm{E}_{\bmxi}$ denotes expectation with respect to the stationary distribution.)\label{prp:mle:4}
	\end{enumerate}
	Then, with a probability that goes to one as $n\to\infty$, a consistent maximum likelihood estimator $\hat\bmxi_n = (\hat\bmbeta_n,\hat\sigma_n)^\prime$ exists and is unique on any compact subset of $\Xi$ containing  $\bmxi_0$. Moreover,
	\begin{align}
		\sqrt{n}(\hat\bmxi_n - \bmxi_0) \stackrel{\mathcal{D}}{\longrightarrow} \mathcal{N}_q(\bzero,\bmcalI^\Delta(\bmxi_0)^{-1}) \label{eq:asympnorm}
	\end{align}
	as $n\to\infty$, where $\bmcalI^\Delta(\bmxi_0)$ can be consistently estimated by $\bmcalI^\Delta_n(\hat\bmxi_n)$ with
	\begin{align}
		\label{eq:infest}
		\bmcalI^\Delta_n(\bmxi) = \frac{1}{n} \sum_{i=1}^n \partial_{\bmxi} \log p_\Delta (\Theta_{i\Delta};\bmxi\mid \Theta_{(i-1)\Delta}) (\partial_{\bmxi} \log p_\Delta (\Theta_{i\Delta};\bmxi\mid \Theta_{(i-1)\Delta)}))^\prime.
	\end{align}
	Finally, let \smash{$\bmxi^\prime = (\bmxi^{(1)\prime},\bmxi^{(2)\prime})$}, where \smash{$\bmxi^{(1)}$} is $q_1$-dimensional and \smash{$\bmxi^{(2)}$} is $q_2$-dimensional ($q_1+q_2=q$). Let \smash{$Q_n \defin \sup_{\bmxi\in\Xi_0} L_n(\bmxi)\big/ \sup_{\bmxi\in\Xi} L_n(\bmxi)$}, with \smash{$\Xi_0\defin\{\bmxi\in\Xi:\bmxi^{(2)} = \bmxi^{(2)}_0\}$}. Then
	\begin{align}
		- 2 \log Q_n \stackrel{\cal D}{\longrightarrow} \chi^2_{q_2} \label{eq:Q}
	\end{align}
	as $n\to\infty$, provided that \smash{$\mathcal{H}_0:\bmxi^{(2)} = \bmxi^{(2)}_0$} holds.
\end{theorem}

The condition that the Fisher information \eqref{eq:Fisher} is satisfied if the coordinates of $\partial_{\bmxi} \log p_\Delta (\theta_2;\bmxi\mid \theta_1)$ are linearly independent functions of $(\theta_1,\theta_2)$. Note that the first term in the first $q-1$ coordinates are the score function of the circular \pdf/ $f_{\bmbeta}$. Therefore, if the Fisher information for the model $f_{\bmbeta}$, $\bmbeta \in B$, is positive definite, then $\smash{\bmcalI^\Delta(\bmxi)}$ is positive definite too, unless the derivatives of $\log  \sum_{k\in\mathbb{Z}} \phi_{\sigma^2\Delta} (h_k(\theta_1,\theta_2;\bmbeta))$ depend on the derivatives of $\log f_{\bmbeta}(\theta_2)$ in a rather pathological way.

The likelihood ratio test based on \eqref{eq:Q} rejects \smash{$\mathcal{H}_0:\bmxi^{(2)} = \bmxi^{(2)}_0$} at the asymptotic significance level $\alpha$ when
\begin{align}
	-2\log Q_n>\chi^2_{q_2;1-\alpha}, \label{eq:lrt}
\end{align}
where $\chi^2_{q;\beta}$ denotes the order-$\beta$ quantile of the chi-square distribution with $q$ degrees of freedom. A possible use of this test is testing toroidal Brownian motionness against driftness and diffusionness: $\mathcal{H}_0: \bmbeta=\bmbeta_0$, where $\bmbeta_0$ is such that $f_{\bmbeta}' \equiv 0$ vs. $\mathcal{H}_1: \bmbeta\neq\bmbeta_0$. In diffusion \eqref{eq:diffvm}, this corresponds to testing $\mathcal{H}_0: \kappa=0$ vs. $\mathcal{H}_1: \kappa>0$, while using the bivariate von Mises diffusion stemming from Example \ref{ex:3}, it corresponds to $\mathcal{H}_0: \kappa_1=\kappa_2=0$ vs. $\mathcal{H}_1: \kappa_1>0\text{ or }\kappa_2>0$. In these last two testing problems, the true parameter does not belong to the interior of the parameter space (third assumption in Theorem \ref{prp:mle}). However, through simulations we have corroborated that using the likelihood ratio test \eqref{eq:lrt} still results in tests that respect the usual asymptotic nominal levels (\sect/\ref{sec:val}). A deeper investigation of these and other boundary cases could be approached with the unrestricted maxima strategy of \cite{Feng1992}. Another interesting test is that of stationary independence, which in the bivariate von Mises diffusion is $\mathcal{H}_0: \lambda=0$ vs. $\mathcal{H}_1: \lambda\neq0$.

\subsection{Numerical inference validation}
\label{sec:val}

Consider the von Mises diffusion given in Example \ref{ex:1}, where \smash{$\bmxi^\prime=\allowbreak(\bmxi^{(1)\prime},\xi^{(2)})$} with \smash{$\bmxi^{(1)\prime}=(\mu,\kappa)$} and \smash{$\xi^{(2)}=\sigma$}. We address next an empirical validation of the inferential results in Theorem \ref{prp:mle} for $\mathcal{H}_0:\bmxi^{(1)}=\bmxi_0^{(1)}$. For that, we simulated $M=10,\!000$ discretized trajectories \smash{$\{\Theta_{i\Delta}^{(j)}\}_{i=0}^n$} from \eqref{eq:diffvm} with $n=50,200$ and $\Delta=0.5$, where $j=1,\ldots,M$. The data generation process (DGP) parameters are $\bmxi_{k}=(0, \kappa_k, 1/(2\pi))^\prime$ with $\kappa_k=1+0.10k$, $k=0,1,2$. The initial \smash{$\Theta_0^{(j)}$'s} were drawn from the stationary distribution $\mathrm{vM}(0,\kappa_k)$. Simulation from \eqref{eq:diffcirc} is trivial and exact from \eqref{eq:sol}; in particular, it does not require an Euler-like discretization. The maximum likelihood estimators $\hat{\bmxi}_n$ were computed by numerically maximizing \eqref{eq:ll}, using the stationary maximum likelihood estimates \cite[\sect/3.1]{Garcia-Portugues2019} as starting values. $\bmcalI^\Delta(\bmxi)$ was approximated by $M$ Monte Carlo samples.

\begin{figure}[h!]
	\centering
	\includegraphics[height=0.415\textheight]{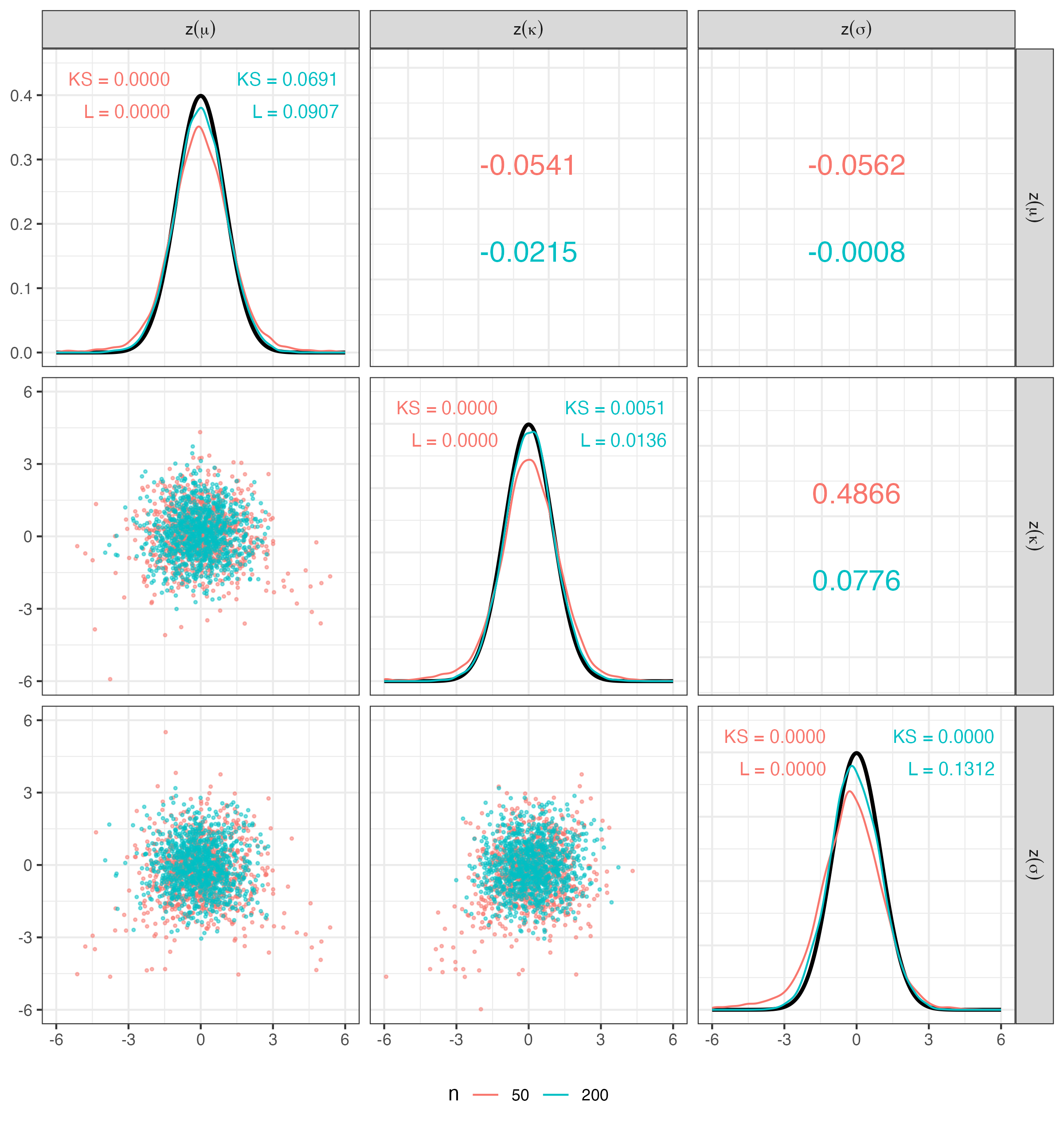}\includegraphics[height=0.415\textheight]{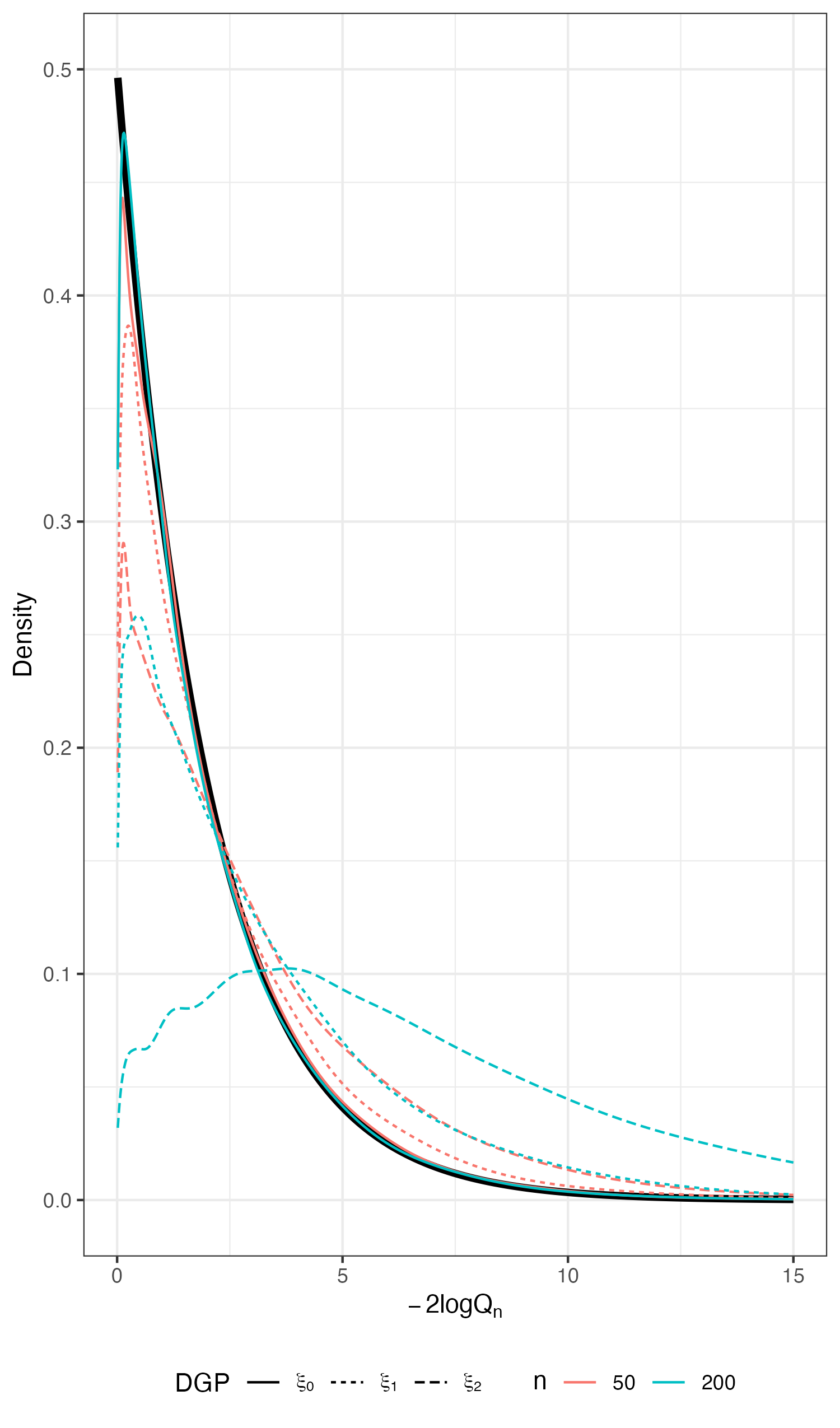}
	\caption{\small Empirical validation of Theorem \ref{prp:mle} for $n=50$ (red) and $n=200$ (turquoise). The left plot shows a diagnostic on the multivariate normality of $\smash{\bmz^{(j)}\defin\sqrt{n}\bmcalI^\Delta(\bmxi_0)^{1/2}(\hat\bmxi_n^{(j)} - \bmxi_0)}$, $j=1,\ldots,M$, with \smash{$\hat\bmxi_n^{(j)}$} estimated with samples from the \dgp/ with $\bmxi=\bmxi_0$. The upper and lower panels show correlations and scatterplots, respectively, while central panels give the standard normal \pdf/ (black), marginal \kdes/, and $p$-values of normality tests. The right plot displays the \pdf/ of the $\chi^2_{2}$ distribution (black) and the \kdes/ of $-2\log Q_n$ under $\smash{\mathcal{H}_0:\bmxi^{(1)}=\bmxi_0^{(1)}}$ and $\smash{\mathcal{H}_1:\bmxi^{(1)}=\bmxi_k^{(1)}}$, $k=1,2$.}
	\label{fig:3}
\end{figure}

The left plot in Figure \ref{fig:3} shows the pairwise scatterplots of \smash{$\bmz^{(j)}\defin\sqrt{n}\bmcalI^\Delta(\bmxi_0)^{1/2}(\hat\bmxi_n^{(j)} - \bmxi_0)$} (lower panels) for a subset of $2,\!000$ observations. The only \dgp/ considered is that with $\bmxi_0$. For $n=50$ (red), non-normal dispersion patterns are evident, especially in the $\kappa$ and $\sigma$ components. This is also reflected in their correlations (upper panels). Also, the marginal kernel density estimators \abbr{(kdes)} signal a deviation from the expected $\phi$ density (black), aligning with the $p$-values of the Kolmogorov--Smirnov (KS) test for $\mathcal{N}(0,1)$ and the Lilliefors (L) test. Correlations and tests were computed from the whole sample $\{\bmz^{(j)}\}_{j=1}^M$. When $n$ increases to $n=200$ (turquoise), the previous metrics improve and convergence towards a $\mathcal{N}_3(\bzero,\bmI_3)$ is evidenced, thus validating numerically \eqref{eq:asympnorm}. The bandwidths of the \kdes/ were chosen with the rule-of-thumb selector.

The convergence in \eqref{eq:Q} under \smash{$\mathcal{H}_0:\bmxi^{(1)}=\bmxi_0^{(1)}$} is neatly corroborated in the right plot in Figure \ref{fig:3} with the close match between the \kdes/ of \smash{$\{-2\log Q_n^{(j)}\}_{j=1}^M$} and the \pdf/ of the $\chi^2_2$ distribution (black). The $p$-values of the KS test for $\chi^2_2$ are $0.0159$ ($n=50$) and $0.1049$ ($n=200$). As expected, under \smash{$\mathcal{H}_1:\bmxi^{(1)}=\bmxi_k^{(1)}$}, $k=1,2$, the \kdes/ diverge from the null $\chi^2_2$ \pdf/ at different paces controlled by the strength $k$ of the deviation (line types) and the sample size $n$ (colours), even though the divergence from $\mathcal{H}_0$ only happens in the $\kappa$ component. The \kdes/ in the plot were obtained through log-transformation and rule-of-thumb bandwidths.

A final validation concerns the empirical rejection rates of \eqref{eq:lrt} under \smash{$\mathcal{H}_0:\bmxi^{(1)}=\bmxi_0^{(1)}$}. For the asymptotic significance levels $\alpha=0.10,0.05,0.01$, these were: $0.1053$, $0.0512$, and $0.0109$ ($n=50$); $0.1014$, $0.0502$, and $0.0105$ ($n=200$). When testing the boundary hypothesis $\mathcal{H}_0:\kappa=0$ for the different \dgps/ with $\bmxi=(0,0,1/(2\pi))^\prime$, the empirical rejection rates were: $0.0929$, $0.0461$, and $0.0100$ ($n=50$); $0.0965$, $0.0477$, and $0.0092$ ($n=200$). Except $0.0929$, all the rejection rates are inside the asymptotic $95\%$ confidence interval for $\alpha$.

\subsection{\texorpdfstring{Test of linear hypotheses for $k$ groups of diffusions}{Test of linear hypotheses for k groups of diffusions}}
\label{sec:anova}

Consider $k$ circular diffusions \eqref{eq:diffcirc} with the same underlying parametric form for the \pdf/ $f_{\bmbeta}$, $\bmbeta \in B \subseteq \mathbb{R}^{q-1}$, but with possibly different parameter values. Let there be $N_j$ independent and identically distributed copies of each diffusion, with $j=1,\ldots,k$. The following uncoupled system of \sdes/ arises:
\begin{align}
	\rd \Theta_t^{(i,j)} = -\frac{\sigma_j^2f_{\bmbeta_j}'\big(\Theta_t^{(i,j)}\big)}{2f_{\bmbeta_j}\big(\Theta_t^{(i,j)}\big)^3}\rd t+\frac{\sigma_j}{f_{\bmbeta_j}\big(\Theta_t^{(i,j)}\big)}\rd W_t^{(i,j)}, \quad t\in[0,T_j],\label{eq:anova}
\end{align}
where $i=1,\ldots,N_j$, $\bmbeta_j\in B$, and $\bmxi_j^\prime=(\bmbeta_j^\prime,\sigma_j)^\prime\in B\times\mathbb{R}^+$ collects the parameters of the $j$th diffusion. An instance of \eqref{eq:anova} is used in \sect/\ref{sec:ants}, where \smash{$\{\Theta_t^{(i,j)}\}$} represents the angles of movement of the $i$th ant in the $j$th group during an experiment lasting time $T_j$. %

Given the discrete time observations \smash{$\{\Theta_{\nu\Delta_j}^{(i,j)}\}_{\nu,i,j=0,1,1}^{n_j,N_j,k}$}, the log-likelihood function of the vector of parameters of \eqref{eq:anova}, \smash{$\bmxi=(\bmxi_1^\prime,\ldots,\bmxi_k^\prime)^\prime\in\Xi=(B\times\mathbb{R}^+)^k$}, is
\begin{align}
	\ell_{\bmn,\bmN}\big(\bmxi; \{\Theta_{\nu\Delta_j}^{(i,j)}\}_{\nu,i,j=0,1,1}^{n_j,N_j,k}\big) = \sum_{j=1}^k\sum_{i=1}^{N_j} \ell_{n_j}\big(\bmxi_j; \{\Theta_{\nu\Delta_j}^{(i,j)}\}_{\nu=0}^{n_j}\big), \label{eq:llanova}
\end{align}
where $\bmn \defin (n_1, \ldots, n_k)$ and $\bmN \defin (N_1, \ldots, N_k)$, and where $\ell_{n_j}$ is given by \eqref{eq:ll}.

We first give a result on the likelihood ratio test for general linear hypotheses for $k$ groups of diffusions.

\begin{theorem} \label{prp:anova}
	Assume conditions \ref{prp:mle:1}--\ref{prp:mle:3} from Theorem \ref{prp:mle} hold, and that condition \ref{prp:mle:4} holds for each $\bmcalI^{\Delta_j}(\bmxi)$, $j=1,\ldots,k$. Consider a subset of unrestricted parameters $A \subseteq \mathbb{R}^{\tilde q}$ ($\tilde{q} < kq$) and a $kq\times\tilde{q}$-matrix $\bf M$ of full rank $\tilde q$ such that $\Xi_0 \defin \bmM A \subset \Xi$, where the notation $\bmM A = \{ \bmM \bma  \mid  \bma \in A \}$ is used. Let \smash{$Q_{\bmn,\bmN} \defin \sup_{\bmxi\in\Xi_0} L_{\bmn,\bmN}(\bmxi)\big/ \sup_{\bmxi\in\Xi} L_{\bmn,\bmN}(\bmxi)$}, with $L_{\bmn,\bmN} \defin \exp(\ell_{\bmn,\bmN})$. Then, provided that the hypothesis \smash{$\mathcal{H}_0:\bmxi\in \Xi_0$} holds,
	\begin{align}
		- 2 \log Q_{\bmn,\bmN} \stackrel{\cal D}{\longrightarrow} \chi^2_{kq - \tilde{q}} \label{eq:Qlinear}
	\end{align}
	as $n_1N_1,\ldots,n_kN_k\to\infty$ with $n_jN_j/(n_1N_1) = O(1)$ for $j=1,\ldots,k $.
\end{theorem}

By essentially the same proof, a similar more general result can be proved for the likelihood ratio test for one linear hypothesis under another less restrictive linear hypothesis.

An important particular case is testing the homogeneity of subsets of parameters of $\bmxi_j$, $j=1,\ldots,k$, across the $k$ groups. Let \smash{$\bmxi_j^\prime =(\bmxi_j^{(1)\prime},\bmxi_j^{(2)\prime})$}, where \smash{$\bmxi_j^{(1)}$} is $q_1$-dimensional and \smash{$\bmxi_j^{(2)}$} is $q_2$-dimensional ($q_1+q_2=q$), for all $j=1,\ldots,k$. The hypothesis \smash{$\mathcal{H}_0:\bmxi_1^{(2)}=\cdots=\bmxi_k^{(2)}$} is of the type considered in Theorem \ref{prp:anova} with
\smash{$\Xi_0\defin\{\bmxi\in\Xi:\bmxi_1^{(2)}=\cdots=\bmxi_k^{(2)}\}$} and $\tilde q = kq_1+q_2$. Therefore,
\begin{align}
	- 2 \log Q_{\bmn,\bmN} \stackrel{\cal D}{\longrightarrow} \chi^2_{(k-1)q_2}. \label{eq:Qanova}
\end{align}
As an example, consider the von Mises diffusion with two groups. Here $\bmxi_j = (\mu_j, \kappa_j, \sigma_j)^\prime \in [0,2\pi)\times(0,\infty)^2$, $j=1,2$. Suppose we wish to test the hypothesis $\mathcal{H}_0: \sigma_1 = \sigma_2$. The unrestricted parameters are then $(\mu_1,\kappa_1,\sigma_1,\mu_2,\kappa_2)\in A = [0,2\pi)\times(0,\infty)^2\times[0,2\pi)\times(0,\infty)$ and
\[
\bmM = \left( \begin{array}{ccccc}
	1 & 0 & 0 & 0 & 0 \\
	0 & 1 & 0 & 0 & 0 \\
	0 & 0 & 1 & 0 & 0 \\
	0 & 0 & 0 & 1 & 0 \\
	0 & 0 & 0 & 0 & 1 \\
	0 & 0 & 1 & 0 & 0
\end{array}\right).
\]

The likelihood ratio test based on \eqref{eq:Qanova} provides substantial flexibility. Within the von Mises diffusion \eqref{eq:diffvm}, perhaps the most immediate case is the ANOVA-type test of equal stationary means ($\mathcal{H}_0:\mu_1=\cdots=\mu_k$), which can be extended to the test of equal stationary distributions ($\mathcal{H}_0:(\mu_1,\kappa_1)=\cdots=(\mu_k,\kappa_k)$) or to the most stringent test of homogeneous diffusions ($\mathcal{H}_0:(\mu_1,\kappa_1,\sigma_1)=\cdots=(\mu_k,\kappa_k,\sigma_k)$). %
Besides, the likelihood ratio test in Theorem \ref{prp:mle} can be straightforwardly adapted to test \smash{$\mathcal{H}_0:\bmxi^{(1)} = \bmxi^{(2)}$} in a two-piece version of the circular diffusion \eqref{eq:diffcirc} featuring a change point at time $T_1$:
\begin{align}
	\rd \Theta_t^{(i)} = -\frac{\sigma_j^2f_{\bmbeta_j}'\big(\Theta_t^{(i)}\big)}{2f_{\bmbeta_j}\big(\Theta_t^{(i)}\big)^3}\rd t+\frac{\sigma_j}{f_{\bmbeta_j}\big(\Theta_t^{(i)}\big)}\rd W_t^{(i)}, \quad j=\begin{cases}
		1, & \text{if } t\in [0, T_1],\\
		2, & \text{if } t\in (T_1,T_2],
	\end{cases}\label{eq:twodiff}
\end{align}
with $0<T_1<T_2$ and $i=1,\ldots,N$.

We conclude the section with a small numerical validation of Theorem \ref{prp:anova} akin to that of \sect/\ref{sec:val}. Let $k=2$, $n_1=n_2=50$, and $\Delta_1=\Delta_2=0.5$. The \dgp/s of both groups are von Mises diffusions. The first has parameters $\bmxi_{1}=(0,1,1/(2\pi))^\prime$. The parameters of the second diffusion differ only in the concentration: $\bmxi_{2,a}=(0, 1+a/10, 1/(2\pi))^\prime$ for $a=0,1,\ldots,5$. We apply the likelihood ratio test for the null hypotheses and sample sizes described in Table \ref{tab:anova} for $M=10,\!000$ discretized trajectories \smash{$\{\Theta_{\ell\Delta_j}^{(i,j)}\}_{\ell,i,j=0,1,1}^{n_j,N_j,k}$}.

As expected, Table \ref{tab:anova} corroborates that: (\textit{i}) the tests maintain the significance level, with the null rejection rates for $(N_1,N_2)=(10,5)$ being within the asymptotic $95\%$ confidence interval for $\alpha$; (\textit{ii}) the tests for $\mathcal{H}_0:\mu_1=\mu_2$ and $\mathcal{H}_0:\sigma_1=\sigma_2$ are `blind' against the considered alternative; (\textit{iii}) the tests for $\mathcal{H}_0:\kappa_1=\kappa_2$, $\mathcal{H}_0:(\kappa_1,\sigma_1)=(\kappa_2,\sigma_2)$ and $\mathcal{H}_0:(\mu_1,\kappa_1)=(\mu_2,\kappa_2)$, and $\mathcal{H}_0:(\mu_1,\kappa_1,\sigma_1)=(\mu_2,\kappa_2,\sigma_2)$ are decreasingly ordered in terms of power; (\textit{iv}) the non-trivial powers increase with $(N_1,N_2)$ ($n_1$ and $n_2$ are fixed).

\section{Diffusion bridge simulation}
\label{sec:bridge}

By virtue of Proposition \ref{prp:Xt2}, simulating a diffusion bridge $\bmTheta_t\mid(\bmTheta_0=\bmtheta_0,\bmTheta_T=\bmtheta_T)$ for \eqref{eq:difftorus} reduces to sampling a certain `winding number' $\bmK_T\in\mathbb{Z}^p$ associated with $\bmtheta_T$, simulating the Brownian bridge $\bmW_t\mid(\bmW_0=\bmx,\bmX_T=\bmy)$ for certain initial and final values $\bmx$ and $\bmy$, and applying a posterior wrapping. The following theorem makes this procedure precise.

\begin{theorem}\label{prp:bridge}
	Let $\{\bmTheta_t\}$ be the toroidal diffusion given by  \eqref{eq:difftorus}, $0<t<T$, and consider $\bmtheta_0, \bmtheta_T\in\mathbb{T}^p$ and $0 < t_1 < \cdots < t_n < T$.
	\begin{enumerate}
		\item Simulate $\bmK_T\in\mathbb{Z}^p$ such that
		\begin{align}
			\mathbb{P}[\bmK_T=\bmk]=
			\frac{\phi_{T\bmSigma}(R(\bmtheta_T)-R(\bmtheta_0)+\bmk)}{\sum_{\bmm\in\mathbb{Z}^p}\phi_{T\bmSigma}(R(\bmtheta_T)-R(\bmtheta_0)+\bmm)},\quad \bmk\in\mathbb{Z}^p.\label{eq:KT}
		\end{align}
		\label{bridge-step1}
		\item Simulate independently $Z_{i,j}\sim \mathcal{N}(0, t_i-t_{i-1})$ for $i=1, \ldots, n+1$ and $j=1,\allowbreak\ldots,p$, with $t_0=0$ and $t_{n+1}=T$, %
		and calculate
		\begin{align}
			U_{i,j} = \sum_{m=1}^i Z_{m,j} -\frac{t_i}{T}\left(\sum_{m=1}^{n+1} Z_{m,j} - y_j \right)
			\label{eq:bridgeU}
		\end{align}
		for $i=1,\ldots,n$, $j=1,\ldots,p$, where $\bmy\defin\bmSigma^{-1/2}(R(\bmtheta_T)-R(\bmtheta_0)+\bmK_T)$. \label{bridge-step2}
		\item Set $\bmB_i \defin R^{-1}(\bmSigma^{1/2} \bmU_i+R(\bmtheta_0))\mod 2\pi$ with $\bmU_i = (U_{i,1}, \ldots,U_{i,p})^\prime$. \label{bridge-step3}
	\end{enumerate}
	Then $(\bmB_1,\ldots,\bmB_n)$ is distributed as $(\bmTheta_{t_1},\ldots,\bmTheta_{t_n})\mid(\bmTheta_0=\bmtheta_0,\bmTheta_T=\bmtheta_T)$.
\end{theorem}

By the Markov property and time-reversibility, the \pdf/ of $\bmTheta_t\mid(\bmTheta_0=\bmtheta_0,\bmTheta_T=\bmtheta_T)$ is
\begin{align*}
	p_{t}(\bmtheta\mid\bmTheta_0=\bmtheta_0,\bmTheta_T=\bmtheta_T)
	=&\;\frac{p_{T-t}(\bmtheta\mid\bmTheta_0=\bmtheta_T)\ p_{t}(\bmtheta\mid\bmTheta_0=\bmtheta_0)}{p_{T}(\bmtheta_T\mid\bmTheta_0=\bmtheta_0)}\frac{f(\bmtheta_T)}{f(\bmtheta)},\quad 0<t<T.
\end{align*}
Figure \ref{fig:4} displays several bridges simulated using Theorem \ref{prp:bridge}. The bridges join two pairs of initial-end points (grey circles and triangles) in three circular/toroidal diffusions with prescribed stationary \pdfs/. As expected, the sample paths show higher/lower volatility at low/high-density regions and honour the periodicity of the support.

\begin{table}[h!]
	\centering
	\begin{tabular}{ll|rrrrrr}
		\toprule
		$(N_1,N_2)$ & Homogeneity & $a=0$ & $a=1$ & $a=2$ & $a=3$ & $a=4$ & $a=5$ \\ \midrule
		$(3,2)$ & Means & $5.37$ & $5.55$ & $5.45$ & $5.34$ & $5.27$ & $5.23$\\
		& Concentrations & $5.38$ & $9.50$ & $21.27$ & $39.65$ & $59.25$ & $75.59$\\
		& Volatilities & $5.12$ & $5.21$ & $5.23$ & $5.20$ & $5.26$ & $5.34$\\
		& Concs. \& volas. & $5.43$ & $8.64$ & $17.19$ & $32.13$ & $51.10$ & $68.79$ \\
		& Stationary distrs. & $5.51$ & $8.47$ & $17.20$ & $31.13$ & $49.43$ & $67.47$\\
		& Diffusions & $5.53$ & $7.83$ & $15.04$ & $27.46$ & $44.90$ & $62.43$\\
		\midrule
		$(10,5)$ & Means & $4.98$ & $5.04$ & $5.12$ & $5.02$ & $4.96$ & $5.11$\\
		& Concentrations & $5.25$ & $18.21$ & $50.44$ & $81.09$ & $96.30$ & $99.42$\\
		& Volatilities & $5.22$ & $5.24$ & $5.27$ & $5.23$ & $5.10$ & $5.13$\\
		& Concs. \& volas. & $5.19$ & $14.07$ & $40.85$ & $74.06$ & $93.03$ & $98.91$ \\
		& Stationary distrs. & $5.17$ & $13.56$ & $40.05$ & $73.03$ & $92.51$ & $98.65$\\
		& Diffusions & $5.22$ & $11.78$ & $35.22$ & $68.15$ & $89.81$ & $98.11$\\
		\bottomrule
	\end{tabular}
	\caption{\small Empirical rejection levels of null hypotheses of homogeneity for $k=2$ groups using the likelihood ratio test based on \eqref{eq:Qanova}. The case $a=0$ represents equal \dgps/; $a>0$ introduces a shift on the concentrations. All tests are performed at the asymptotic significance level $\alpha=0.05$.}
	\label{tab:anova}
\end{table}

\section{Circular jump processes with exact likelihood inference}
\label{sec:jump}

The construction in the previous sections can also be applied to other L\'evy processes. To demonstrate this, we present a particularly useful case based on the Cauchy process.

Let $\{V_t\}$ be a (symmetric) Cauchy process with scale parameter $\sigma$. Then $V_{t_2}-V_{t_1}$ is $\mathrm{C}(0,((t_2-t_1)\sigma)^2)$-distributed for $t_2>t_1>0$, where $\mathrm{C}(\mu,\sigma^2)$ denotes the Cauchy distribution given in Example \ref{ex:2}. As in Section \ref{sec:circ}, let $f:\mathbb{R}\to\mathbb{R}^+$ be a given circular \pdf/, and let $F:\mathbb{R}\to\mathbb{R}$ be the associated circular \cdf/, and define the jump process
\begin{align}
	\Theta_t=F^{-1}(V_t+F(\theta_0))\mod 2\pi.\label{eq:jum1}
\end{align}
It follows by arguments analogous to those given in the diffusion case that the \tpd/ of $\{\Theta_t\}$~is
\begin{align*}
	p_{t}(\theta_2 \mid \theta_1) %
	=&\; \frac{(1-\rho_t(\sigma)^{2})f(\theta_2)}{{1+\rho_t(\sigma)^{2}-2 \rho_t(\sigma) \cos (2\pi (F({\theta_2})-F(\theta_1)))}},
\end{align*}
where $\rho_t(\sigma)\defin e^{-2\pi t \sigma}$. The process is time-reversible and ergodic with stationary \pdf/ $f$. The asymptotic results on likelihood inference in Theorem \ref{prp:mle} hold for the circular jump process under the same conditions as in Theorem \ref{prp:mle}.

An algorithm for simulating a circular jump process bridge similar to the one in Theorem \ref{prp:bridge} can be obtained by using that a Cauchy process is a stochastic time transformation of a Wiener process, see \citet[\sect/1.3]{Applebaum2009}. Let $\{\tilde{W}_t\}$ be a Wiener process with infinitesimal variance $2\sigma^2$, and $\{S_t\}$ an independent L\'evy subordinator, i.e., $\{S_t\}$ is a non-negative, increasing L\'evy process, where $S_t$ is L\'evy-distributed with scale parameter $\beta = t^2/2$.%
Then $V_t = \tilde{W}_{S_t}$ is a Cauchy process with scale parameter $\sigma$. If $X \sim \mathcal{N}(0,\beta^{-1})$, then $X^{-2}$ is L\'evy distributed with scale parameter $\beta$.

\begin{theorem}\label{prp:jumpbridge}
	Let $\{\Theta_t\}$ be the circular jump process given by \eqref{eq:jum1}, $0<t<T$, and consider $\theta_0, \theta_T\in\mathbb{T}^1$ and $0 = t_0 < t_1 < \cdots < t_n < t_{n+1} = T$.
	\begin{enumerate}
		\item Simulate $K_T\in\mathbb{Z}$ such that \label{jumpbridge-step1}
		\begin{align}
			\mathbb{P}[K_T=k]=
			\frac{T\sigma[1+\rho_T(\sigma)^2 -2\rho_T(\sigma)\cos(2\pi(F(\theta_T)-F(\theta_0)))]}{\pi(1-\rho_T(\sigma)^2)[(F(\theta_T)-F(\theta_0)+k)^2+(T\sigma)^2]},\quad k\in\mathbb{Z}.
		\end{align}
		\item Simulate independently the vector $(A_1,\ldots,A_{n+1})$ with joint \pdf/
		\begin{align}
			f_A(a_1,\ldots,a_{n+1}) %
			=&\;\frac{\phi_{2\sigma^2(a_1^{-2}+\cdots+a_{n+1}^{-2})}(y)}{f_{\mathrm{C}}(y;0,(T\sigma)^2)}\prod_{i=1}^{n+1} \phi_{2\Delta_i^{-2}}(a_i),\label{eq:fA}
		\end{align}
		where $y\defin F(\theta_T)-F(\theta_0)+K_T$, $\Delta_i \defin t_i-t_{i-1}$, and $f_{\mathrm{C}}$ denotes the \pdf/ of the Cauchy distribution, see Example \ref{ex:2}. \label{jumpbridge-step2}
		\item Conditionally on $A_1,\ldots,A_{n+1}$, simulate independently $Z_j\sim \mathcal{N}(0, 2\sigma^2 A_j^{-2})$ for $j=1, \ldots, n+1$ and calculate
		\begin{align*}
			U_i = \sum_{j=1}^i Z_j -\frac{S_{t_i}}{S_{t_{n+1}}}\left(\sum_{j=1}^{n+1} Z_j - y \right), %
		\end{align*}
		$i=1,\ldots,n$, where $S_{t_i} = A_1^{-2}+\cdots+A_i^{-2}$. \label{jumpbridge-step3}
		\item Set $B_i \defin F^{-1}(U_i+F(\theta_0))\mod 2\pi$, $i=1,\ldots,n$. \label{jumpbridge-step4}
	\end{enumerate}
	Then $(B_1,\ldots,B_n)$ is distributed as $(\Theta_{t_1},\ldots,\Theta_{t_n})\mid (\Theta_0=\theta_0,\Theta_T=\theta_T)$.
\end{theorem}

The random vector $(A_1,\ldots,A_{n+1})$ in Step 2 can, for instance, be simulated by an acceptance-rejection approach. A simple proposal density is the independent normal factor in \eqref{eq:fA}, with $M_A\defin [\pi(y^2+(T\sigma)^2)][T\sigma\sqrt{2\pi e} |y|]^{-1}$ being a (tight) bound for the first factor. Thus the efficiency of this acceptance-rejection algorithm is $1/M_A$. Figure \ref{fig:4} displays several jump-process bridge realizations simulated using Theorem \ref{prp:jumpbridge}.

Note that Steps \ref{jumpbridge-step2} and \ref{jumpbridge-step3} in Theorem \ref{prp:jumpbridge} give an algorithm for simulating a discretized path $(U_1,\ldots,U_n)$ from the Cauchy bridge process $(V_{t_1},\ldots,V_{t_n})|(V_0=0,V_T=y)$.

\begin{figure}[h!]
	\centering
	\includegraphics[height=0.235\textheight]{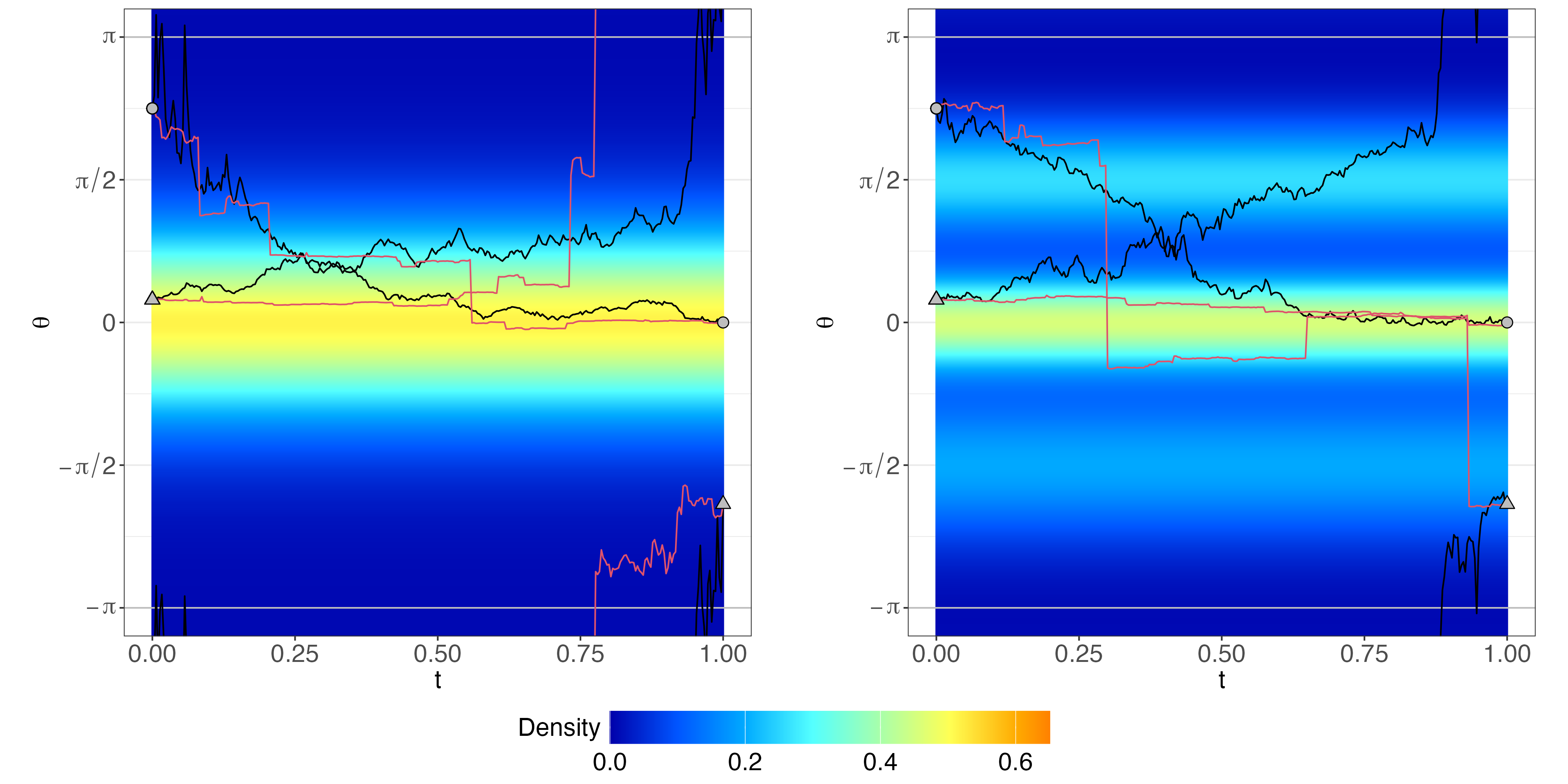}\hspace*{-0.25cm}\includegraphics[height=0.235\textheight]{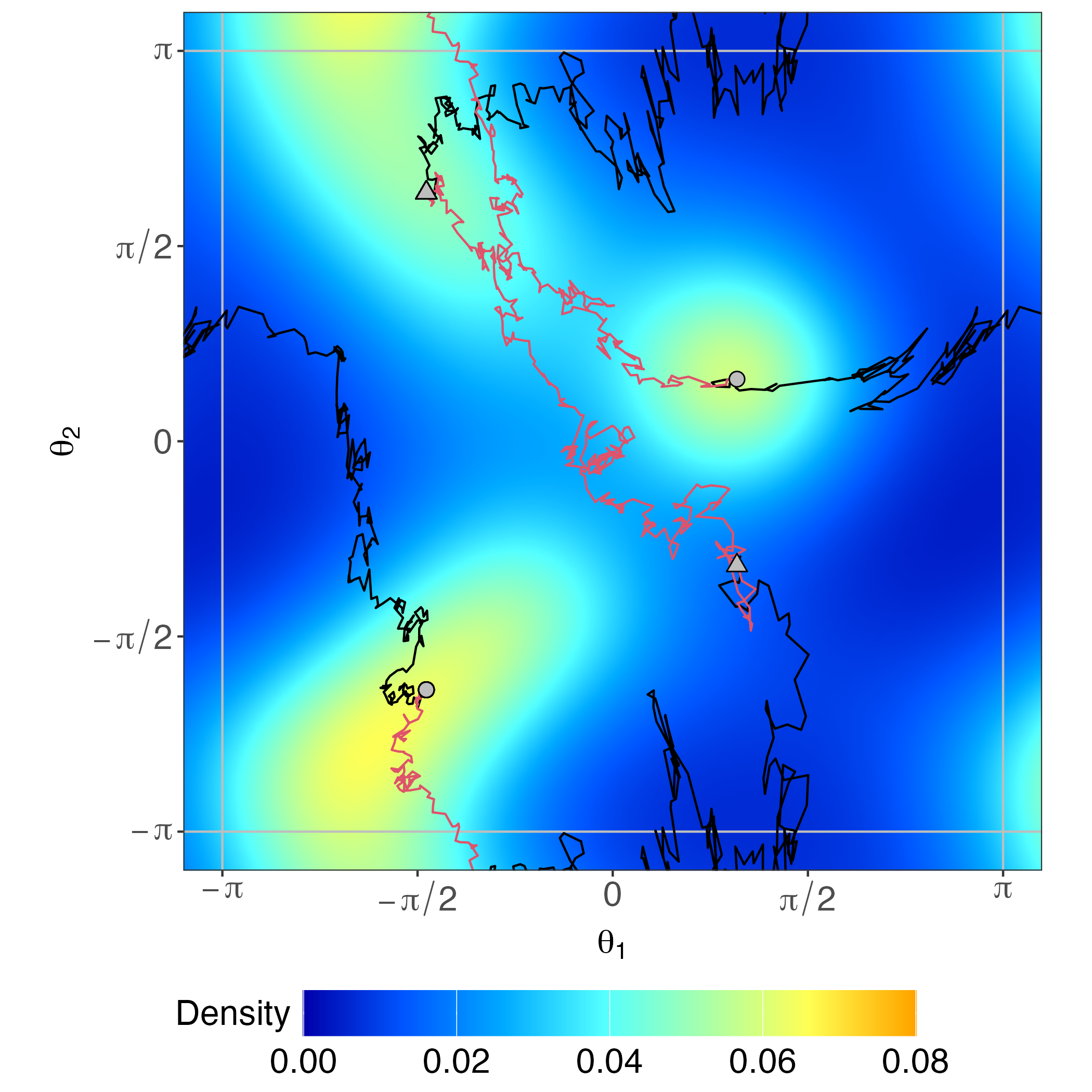}
	\caption{\small Simulated bridges in circular and toroidal processes, with overlaid stationary densities. The left and central plots show the $\mathrm{vM}(0,2)$ and $0.4\mathrm{vM}(0,8)+0.3\mathrm{vM}(-\pi/2,3)+0.3\mathrm{vM}(\pi/2,5)$ \pdfs/, respectively (see Figure \ref{fig:1}). Here, black and red paths stand for diffusion and jump-process bridges, respectively. The right plot shows the $0.4\mathrm{BvM}(-1.5,2,1,1,-0.5)+0.4\mathrm{BvM}(-1.5,-2,1,1.5,1)+0.2\mathrm{BvM}(1,0.5,2,2,0)$ \pdf/ (see Figure \ref{fig:2}). Here, all the paths are diffusion bridges. In all the plots, $\sigma=0.15$ ($\bmSigma=\sigma^2\bmI_2$) and $T=1$.}
	\label{fig:4}
\end{figure}

\section{Biological applications}
\label{sec:data}

\subsection{Homogeneity of ant movements}
\label{sec:ants}

In ant colonies, the `brood-to-worker ratio' is defined as the number of brood items (larvae) divided by the number of workers (adult non-reproductive ants). This metric informs on the quality of care provided to the brood in a colony \citep{Cassill1999}, which has been established as a relevant shaper of the offspring size \citep[see, e.g.,][]{Purcell2012}.

\begin{figure}[h!]
	\centering
	\includegraphics[width=0.34\textwidth, align=c,clip,trim={0cm 1cm 0cm 1cm}]{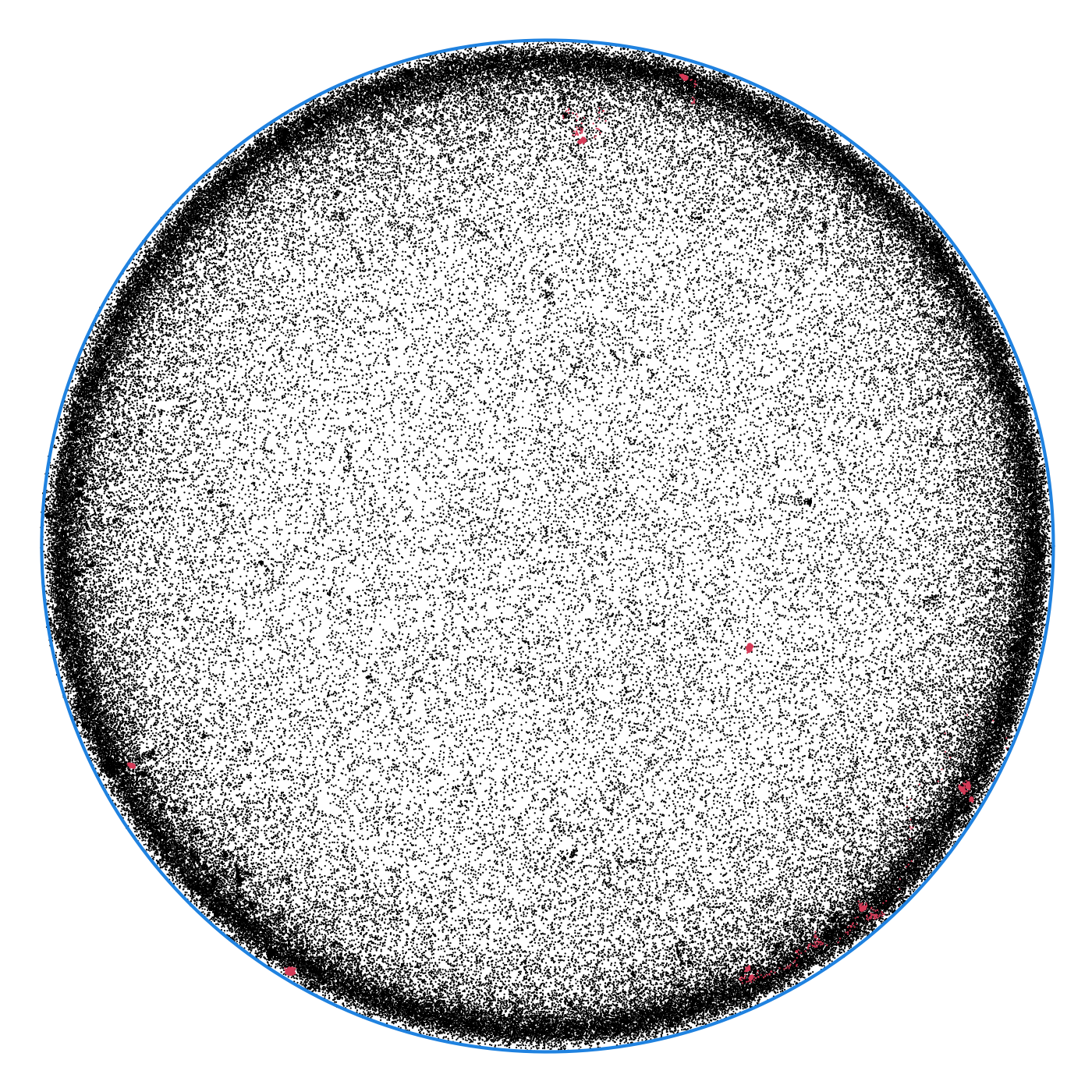}\hspace*{-0.1cm}\includegraphics[width=0.34\textwidth, align=c,clip,trim={0cm 1cm 0cm 1cm}]{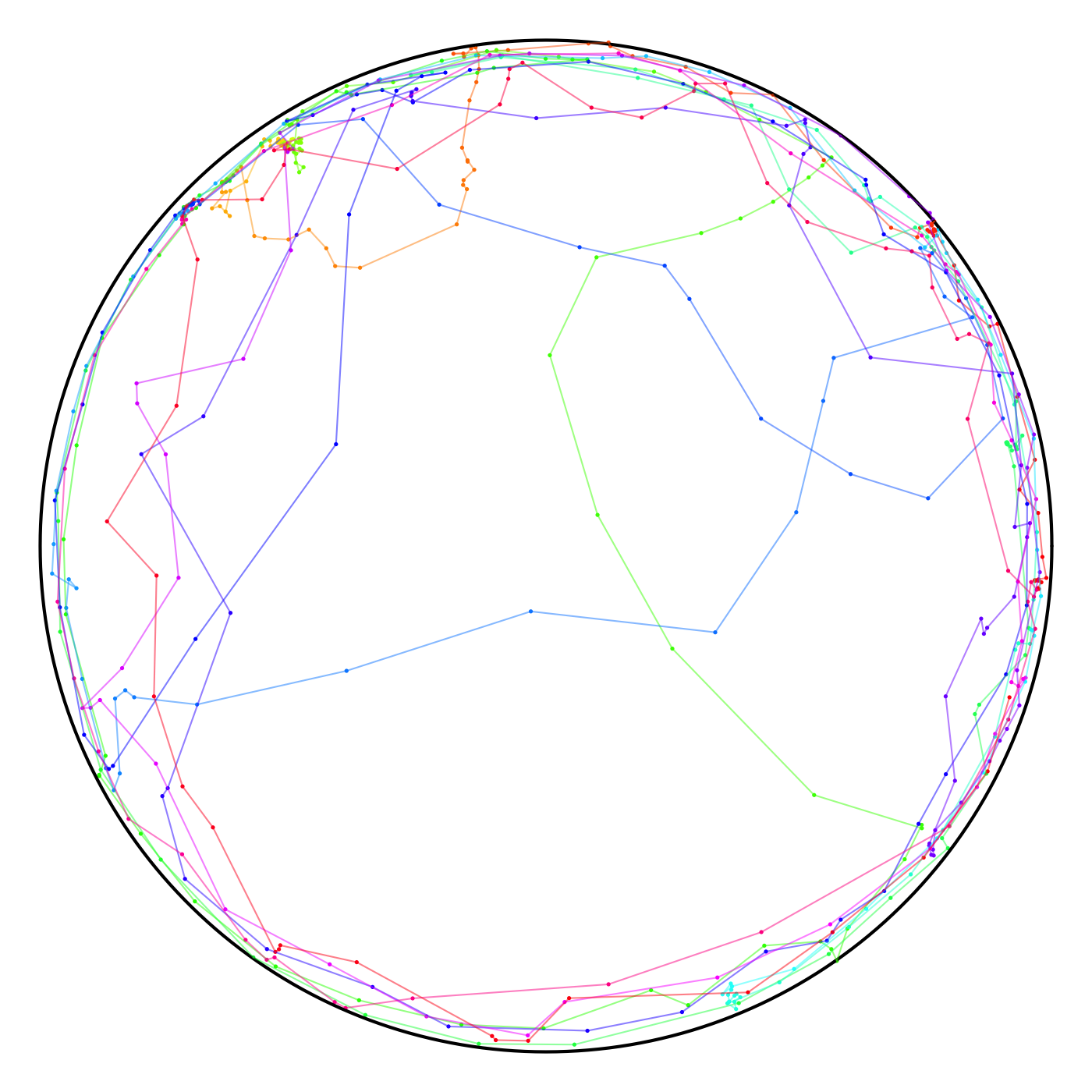}\includegraphics[width=0.32\textwidth, align=c,clip,trim={0cm 0cm 0cm 0cm}]{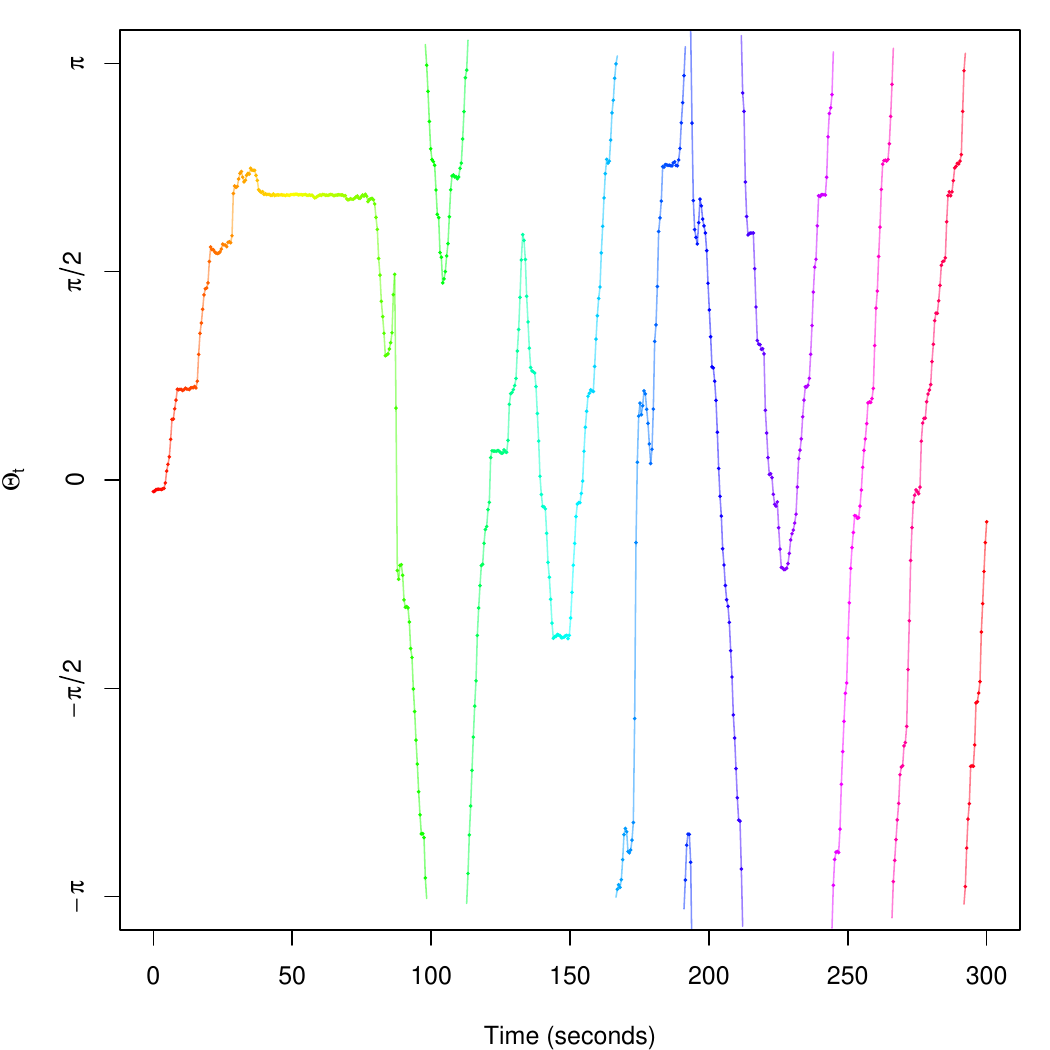}%
	\vspace*{-0.25cm}
	\caption{\small Left: Recorded coordinates of the $249$ ants analysed along the first five minutes after being introduced in the circular arena (blue border). The red points indicate the eight ants that did not move in that time period. Centre: Recorded coordinates $\{(X_t,Y_t)\}$ of the track of a single ant. Right: Angles $\{\Theta_t\}$ of the ant with respect to the centre of the circular arena. The rainbow colour palette indicates the evolution of time and is shared between the last plots.}
	\label{fig:5}
\end{figure}

\cite{SanmartinVillar2021} studied the effect of the brood-to-worker ratio during larval development in the behaviour of workers in the adult stage. Their study on \emph{Lasius niger} ants involved raising ant larvae in nests with two brood-to-worker ratios: 3:1 (high brood treatment) and 1:3 (low brood treatment). Each of the callow workers that eclosed from the larvae was first exposed to the same social treatment: cohabitation with 22 mature workers in a Petri dish for 48 hours. After this period, the behavioural assays of the experiment took place. Each callow worker was put alone in a circular arena ($\varnothing=5.8\text{cm}$, $\text{height} = 1\text{cm}$) covered with a glass lid and with walls coated with a slippery substance. The ant was recorded on video and then its track coordinates were obtained with motion tracking software at a rate of $2.08$ observations per second. %
Figure \ref{fig:5} (centre) shows the coordinates of the track of an ant in the circular arena.

As shown in Figure \ref{fig:5} (left and centre), ants tend to move on the periphery of the circular arena in a wall-following behaviour known as thigmotaxis. Thigmotaxis is a strategy used by ants to better orient themselves in the environment \citep{Pratt2001,Dussutour2005}. Due to this characteristic movement pattern, the sequence of ant coordinates in the arena $\{(X_t,Y_t)\}$ is well summarized by the sequence of angles $\{\Theta_t\defin\mathrm{atan2}(Y_t,X_t)\}$ with respect to the centre of the arena, assumed as the origin.

We use von Mises diffusions to model $\{\Theta_t\}$ and conduct tests of homogeneity of the model parameters of $k$ groups of diffusions (Theorem \ref{prp:anova}) for different factors derived from \cite{SanmartinVillar2021}'s experiment: `brood treatment' ($k=2$ treatments: high and low, previously explained), `colony' (the adults and larvae used in the experiment were extracted from $k=3$ different colonies: A, B, and C), `size' ($k=2$ groups defined by the median) and `larval development' ($k=4$ groups induced by quartiles). `Larval development' measured the number of days between egg and adult metamorphosis of a worker. Ants that undergo longer larval development are expected to reach adulthood with improved fitness levels. Splits of the ant size into more than two groups were highly unbalanced due to size repetitions, and thus discarded. The homogeneity tests serve to investigate whether the exploration behaviour of ants depends on the aforementioned factors.

We analyse the movements of the ants for the first five minutes of the first assay of the experiment, as this period could be assumed as the exploration stage of the ant within the new environment \citep[e.g.,][]{Reale2007}. Due to limitations of the tracking software, some ants have missing coordinates in their tracks. This occurred, for example, when an ant was shadowed by the walls of a Petri dish that is oblique to the lens of the recording camera. To avoid artefacts in the estimation of diffusion models, in our analysis we excluded ants with more than $5\%$ missing data and imputed the missing coordinates of the remaining ants by linear interpolation on the coordinates $\{(X_t,Y_t)\}$. Eight ants were found to be (virtually) immobile during the considered time (red points in Figure \ref{fig:5}) and therefore were excluded from the main analysis. The resulting preprocessed data are the discretized trajectories \smash{$\big\{\Theta_{\ell\Delta}^{(i,j)}\big\}_{\ell,i,j=0,1,1}^{n,N_j,k}$}, with $n=626$ and $\Delta=0.48\text{s}$, and a total of $N_1+\cdots+N_k=241$ ants.

\begin{table}[h!]
	\centering
	\begin{tabular}%
		{L{3cm}|C{2cm}C{2cm}C{2cm}C{2cm}}
		\toprule
		Homogeneity & Brood & Colony & Size & Larval dev. \\ \midrule
		Means & $1.0000$ & $0.8363$ & $1.0000$ & $4.9\cdot 10^{-5}$\\
		Concentrations & $0.3376$ & $1.0000$ & $1.0000$ & $0.0264$\\
		Volatilities & $0.9096$ & $1.0000$ & $1.0000$ & $0.1320$\\
		Stationary distrs. & $0.1097$ & $1.0000$ & $1.0000$ & $1.2\cdot 10^{-4}$\\
		Concs. and volas. & $0.3555$ & $0.9096$ & $1.0000$ & $5.8\cdot 10^{-11}$\\
		Diffusions & $0.3376$ & $1.0000$ & $1.0000$ & $2.0\cdot 10^{-11}$\\
		\bottomrule
	\end{tabular}
	\caption{\small Corrected $p$-values of the tests of homogeneity of von Mises diffusions, for several types of hypotheses on $(\mu_j,\kappa_j,\sigma_j)$, $j=1,\ldots,k$, (rows) and different factors considered (columns). Correction is done with the false discovery rate method \citep{Benjamini2001}.}
	\label{tab:ants}
\end{table}

Table \ref{tab:ants} gives the corrected $p$-values of the $k$-sample tests for different homogeneity hypotheses. The groups of diffusions generated by the brood treatment ($N_{\mathrm{high}}=113$, $N_{\mathrm{low}}=128$) are not seen to be significantly different (lowest $p$-value: $0.1097$). Therefore, no evidence of the impact of the brood-to-worker ratio in the exploration behaviour of ants, as proxyed through the von Mises diffusion model fitted to the angular movement, is found. This conclusion is in line with those in \cite{SanmartinVillar2021}, where the brood treatment was found to not significantly affect ant behaviour metrics obtained from the coordinates of the tracks. The colony of origin ($N_{\mathrm{A}}=89$, $N_{\mathrm{B}}=68$, $N_{\mathrm{C}}=84$) and size ($N_1=100$, $N_2=136$; five sizes are missing) do not significantly affect the exploration behaviour of ants either. Finally, the `larval development' of the ants is found to affect the movement of ants, with the differences between groups being mostly attributed to different mean and concentration parameters.

The inclusion of immobile ants in the tests did not affect their decisions for `brood' or `size'. For `colony', however, this led to the rejection of several null hypotheses driven by the differences in the means ($p$-value: $3.1\cdot10^{-8}$), which is explainable by the fact that half of the immobile ants belong to Colony A. The effect of the inclusion of immobile ants in `larval development' was a reduction in its $p$-values, with those for the homogeneity of concentrations and volatilities being $9.3\cdot 10^{-5}$ and $4.9\cdot10^{-5}$, respectively.

To control the false discovery rate in the $48$ tests conducted in the analysis, the method by \cite{Benjamini2001} was used to correct and report all the resulting $p$-values.

\subsection{Protein structure bridging}
\label{sec:prots}

The toroidal diffusion \eqref{eq:difftorus2} and its exact bridge sampling can be applied to construct data-driven bridges between the backbones of two protein structures. The three-dimensional coordinates of the atoms in the main chain of a protein $\mathsf{p}$ formed by $N$ amino acids can essentially be parameterized as a sequence of pairs of dihedral angles $\{(\phi_i,\psi_i)\}_{i=1}^{N}\subset\mathbb{T}^{2}$. These angles describe the (sequential) relative positions between the planes formed by the atoms C$_{\alpha,i-1}$--C'$_{i-1}$--N$_{i}$--C$_{\alpha,i}$ and C$_{\alpha,i}$--C'$_{i}$--N$_{i+1}$--C$_{\alpha,i+1}$ of the polypeptide main chain formed by the repetition of the N--C$_{\alpha}$--C' block. The $i$th amino acid side-chain is attached to C$_{\alpha,i}$, and $(\phi_i,\psi_i)$ are the rotation angles about the axes defined by the N$_{i}$--C$_{\alpha,i}$ and C$_{\alpha,i}$--C'$_{i}$ bonds, respectively (the angles $\phi_1$ and $\psi_{N}$ are undefined). The bond C'--N is largely fixed due to its partial double bond nature, which generates the planar configurations of atoms. The dihedral-pairs representation is a parsimonious rotation-invariant parametrization of the backbone of a protein that, assuming ideal bond angles and lengths, captures essentially all the interesting variability in protein conformation \citep[e.g.,][p. 173]{Richardson1981}. For two proteins $\mathsf{p}_A$ and $\mathsf{p}_B$ with $N=N_A=N_B$ (possibly different) amino acids, the relation between their three-dimensional main chains (which, in an abuse of notation, are also referred to as $\mathsf{p}_A$ and $\mathsf{p}_B$ in the sequel) is therefore essentially encoded in the tetra of dihedral angles $\{((\phi_{A,i},\psi_{A,i}),(\phi_{B,i},\psi_{B,i}))\}_{i=1}^{N}$.

Calmodulin (an abbreviation for CALcium MODULated proteIN) acts as an intermediary protein that senses calcium levels and relays signals to various calcium-sensitive enzymes, ion channels and other proteins \citep{Dutta2003}. Calmodulin has four Ca$^{2+}$ binding sites, called EF-hands, that, when they are bound to calcium, transform the closed shape of the calcium-free calmodulin into an more open one. This shift of the three-dimensional protein structure is exemplified in the calcium-free calmodulin $\mathsf{p}_A\equiv\texttt{1cfd}$ \citep{Kuboniwa1995}, found in the African clawed frog (\textit{Xenopus laevis}), and the Ca$^{2+}$-calmodulin $\mathsf{p}_B\equiv\texttt{1rfj}$ \citep{Yun2004}, found in the potato (\textit{Solanum tuberosum}). Both proteins are $N=148$ amino acids long, with coincidences in $133$ amino acid sites ($89.86\%$). We intend to simulate protein bridges between (the backbones of) $\mathsf{p}_A$ and $\mathsf{p}_B$ to emulate the transformation of the calmodulin in the presence of calcium and an alteration in its sequence of amino acids.%

Our bridging between proteins $\mathsf{p}_A$ and $\mathsf{p}_B$ is based on three main steps: (\textit{i}) for each amino acid site $i=1,\ldots,N$, learn a stationary density $f_i$ to construct a toroidal diffusion \eqref{eq:difftorus2}; (\textit{ii}) simulate dihedral bridges $\smash{\{(\Phi_i,\Psi_i)_{\nu\Delta}\}_{\nu=0,i=1}^{n,N}}$ such that $(\Phi_i,\Psi_i)_{0}=(\phi_{A,i},\psi_{A,i})$ and $(\Phi_i,\Psi_i)_{n\Delta}=(\phi_{B,i},\psi_{B,i})$,  $i=1,\ldots,N$; (\textit{iii}) invert the dihedral-pairs representation to merge these dihedral bridges into protein bridges between $\mathsf{p}_A$ and $\mathsf{p}_B$. We elaborate on the details of each step next. %

First, to learn a sensible stationary density $f_i$ for the $i$th amino acid site we used the \texttt{top500} dataset, which contains $500$ protein structures that are non-redundant and have been measured with a high precision \citep{Word1999}. For $i=1,\ldots,N$, we fitted a mixture of bivariate von Mises distributions $\hat{f}_i$ using the subset $\texttt{dih}_i$ of dihedral angles in the \texttt{top500} dataset whose amino acid is either $\mathrm{aa}_{A,i}$ or $\mathrm{aa}_{B,i}$, the respective amino acids of $\mathsf{p}_A$ and $\mathsf{p}_B$ in their $i$th sites. The estimate $\hat{f}_i$ captures the particular toroidal density landscape in which the transition from $(\phi_{A,i},\psi_{A,i})$ to $(\phi_{B,i},\psi_{B,i})$ happens, with a possible change of amino acid involved. There are $42$ unique $\texttt{dih}_i$ datasets, and their average size is $9,\!164$ %
dihedral pairs. For each amino acid site, the bivariate von Mises mixtures were fit using Expectation-Maximization (EM) with the following procedure: (a) run EM fits for $m=1,\ldots,20$ mixture components, initializing the algorithm with fast high-concentration Gaussian approximations \citep{Mardia2007} and then using (numerical) maximum likelihood estimation of the bivariate von Mises parameters; (b) choose the number of components $\hat{m}$ that gives the minimum BIC among $20$ randomly-initialized EM fits for each $m$. The obtained $\hat{m}$ range in $[5,19]$. %
In (a), unimodality of the bivariate von Mises components was enforced to avoid identifiability issues. Additionally, $\hat{f}_i$ was smoothed by truncating the largest concentration parameters to five. %
This smoothing is relevant to avoid simulated diffusion bridges with large jumps caused by extreme drifts or volatilities. Figure \ref{fig:7} shows the density contours of $\hat{f}_i$ for the amino acid sites $i=19,41,96,135$, corresponding to the pairs of amino acids $(\mathrm{aa}_{A,i},\mathrm{aa}_{B,i})$ being (Phenylalanine, Phenylalanine), (Glutamine, Glutamine), (Glycine, Glutamine), and (Glutamine, Glutamine), respectively.

Second, for the $i$th amino acid site we sampled diffusion bridges $\smash{\{(\Phi_i,\Psi_i)_{\nu\Delta}\}_{\nu=0,i=1}^{n,N}}$ such that $(\Phi_i,\Psi_i)_{0}=(\phi_{A,i},\psi_{A,i})$ and $(\Phi_i,\Psi_i)_{T}=(\phi_{B,i},\psi_{B,i})$, $i=1,\ldots,N$, for the diffusion \eqref{eq:difftorus2} given by $(1-\alpha)\hat f_i+\alpha/(2\pi)^2$ with a tuning parameter $\alpha\in[0,1]$ that measures how `density-driven' ($\alpha=0$) or `shortest-path driven' ($\alpha=1$) is the diffusion bridge. We set $n=113$ and $\Delta=1/n$ (hence, $T=1$), and $\sigma=1/(2\pi)^2$. Figure \ref{fig:7} shows three simulated bridges between $(\phi_{A,i},\psi_{A,i})$ and $(\phi_{B,i},\psi_{B,i})$ for $i=19,41,96,135$, one for each $\alpha=0.25,0.5,1$. While for $i=19$ the pairs are relatively close and thus the simulated bridges follow a predictable path, for $i=41$ the paths exploit the torus geometry and for $i=96,135$ are affected by the underlying density $\hat{f}_i$, which for $\alpha>0$ significantly bends the Brownian bridges.

\begin{figure}[h!]
	\centering
	\includegraphics[width=\textwidth]{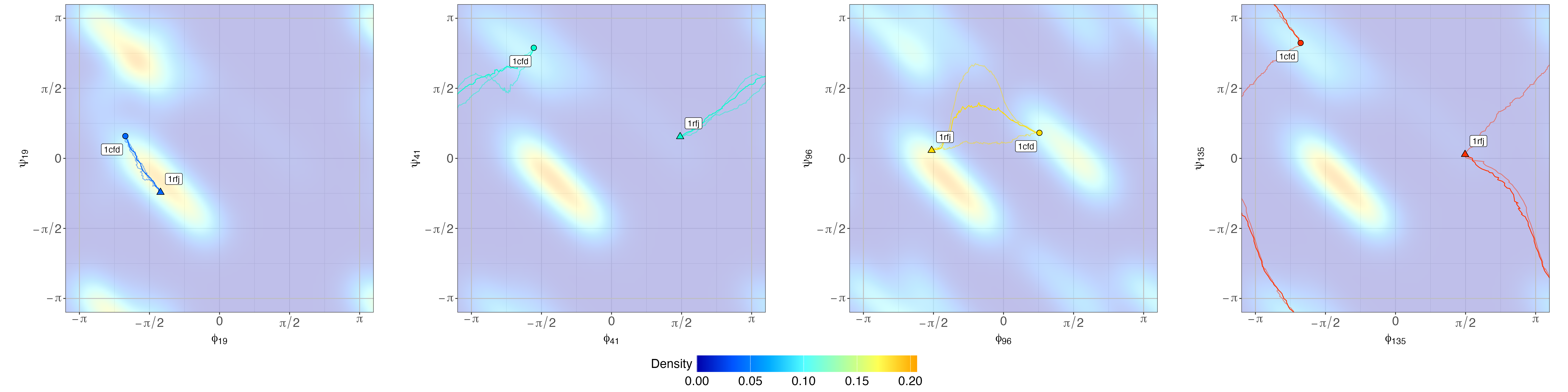}%
	\caption{\small From left to right, the contour plots show the estimated density of the dihedral angle pairs for the amino acids Phenylalanine, Glutamine, Glycine and Glutamine, and Glutamine, respectively, in the \texttt{top500} dataset. These density estimates $\hat{f}_i$ are used to build the toroidal diffusion for the amino acid sites $i=19, 41, 96, 135$ of the proteins \texttt{1cfd} and \texttt{1rfj}. Three simulated diffusion bridges joining the $i$th dihedral angle pairs of these proteins are depicted; the path for $\alpha=0.5$ is shown in solid colours and the two paths for $\alpha=0.25,1$ are shown with transparency.}
	\label{fig:7}
\end{figure}

Third and finally, we inverted the dihedral-pairs representation to obtain the three-dimensional protein backbone structures $\smash{\{\mathsf{P}_{\nu\Delta}\}_{\nu=0}^n}\subset\mathbb{R}^{3}$ from the dihedral bridges $\smash{\{(\Phi_1,\Psi_1,\ldots,\Phi_N,\Psi_N)_{\nu\Delta}\}_{\nu=0}^{n}}\subset\mathbb{T}^{2N}$. To construct $\smash{\{\mathsf{P}_{\nu\Delta}\}_{\nu=0}^n}$, we used first $\mathsf{p}_A$ as structural basis (with its particular amino acid sequence) and modify its dihedral angles $\smash{\{(\phi_{A,i},\psi_{A,i})\}_{i=1}^{N}}$ according to the sampled dihedral bridges to reconstruct the three-dimensional structure, yielding $\smash{\{\mathsf{P}_{A,\nu\Delta}\}_{\nu=0}^n}$. After aligning the three-dimensional structure $\mathsf{p}_B$ to that of $\mathsf{p}_A$, we obtained the time-reverted bridge $\smash{\{\mathsf{P}_{B,(n-\nu)\Delta}\}_{\nu=0}^n}$ analogously. These `forward' and `backward' bridges were merged with the linear interpolation (in $\mathbb{R}^3$) $\mathsf{P}_{\nu\Delta}\defin (1-\nu/n)\mathsf{P}_{A,\nu\Delta}+(\nu/N)\mathsf{P}_{B,\nu\Delta}$ in $\mathbb{R}^3$ to yield the advocated protein bridge. This symmetrization is applied to ensure that $\mathsf{P}_0=\mathsf{p}_A$ and $\mathsf{P}_T=\mathsf{p}_B$, since due to the different amino acid sequences involved, the dihedral-pairs representation inversion yields $\mathsf{P}_{A,T}\neq\mathsf{p}_B$ and $\mathsf{P}_{B,T}\neq\mathsf{p}_A$. Figure \ref{fig:6} shows three simulated protein bridges between $\mathsf{p}_A$ (top left) and $\mathsf{p}_B$ (bottom right), and highlights with coloured spheres the positions of the amino acids whose dihedral bridges are shown~in~Figure~\ref{fig:7}.

The protein bridges shown in Figure \ref{fig:6} give simulated transformations of the closed shape of calcium-free calmodoluin \texttt{1cfd} into the more open shape of the Ca$^{2+}$-calmodulin \texttt{1rfj}. In these simulated bridges, an opening of the protein backbone, much more marked than the open shape of \texttt{1rfj}, is followed by an elongation of the backbone and then a contraction to converge to \texttt{1rfj}. Along this opening and closing of the protein structure, large torsions at certain parts of the main chain take place. For example, between the protein begin and the blue ball (amino acid sites 1--19), five amino acid sites concentrate the large torsions, and between the blue and cyan spheres (20--41), eleven sites 20--41 concentrate the main torsions. The unfolding of the protein backbone is more marked for $\alpha=0.25$ and less for $\alpha=1$.

\begin{figure}[h!]
	\centering
	\includegraphics[width=0.333\textwidth,align=c,clip,trim={0.75cm 10.0cm 6.5cm 10.0cm}]{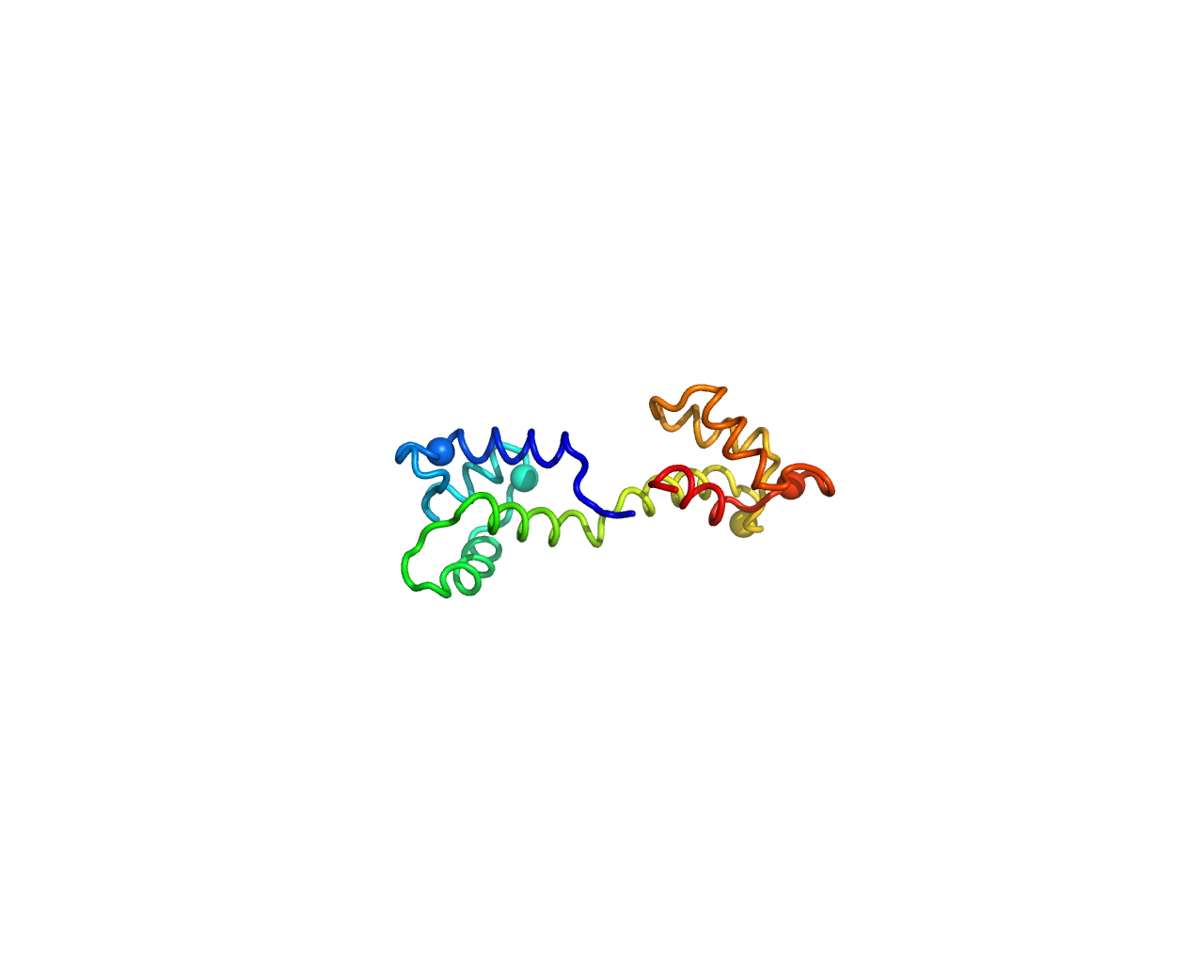}\includegraphics[width=0.333\textwidth,align=c,clip,trim={0.75cm 10.0cm 6.5cm 10.0cm}]{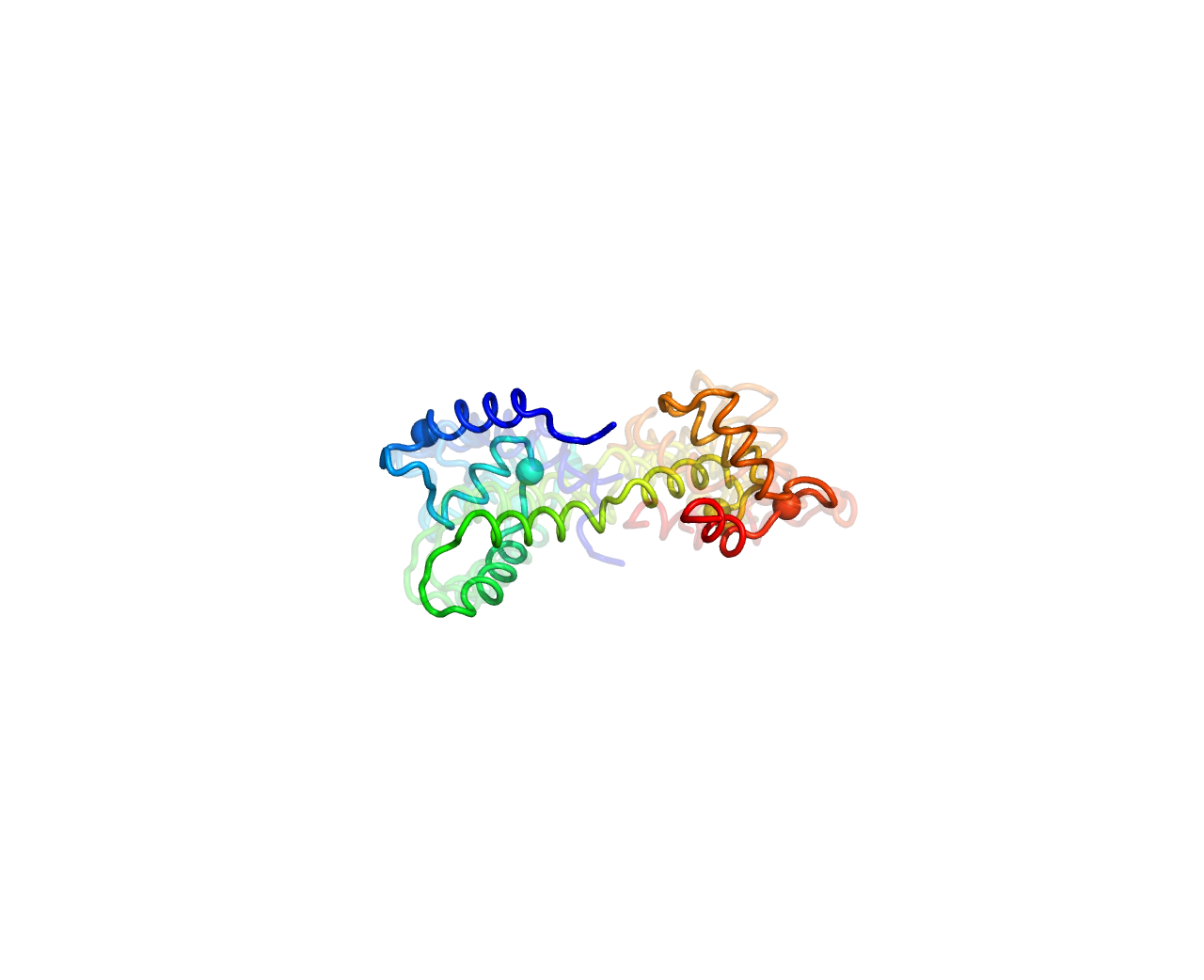}\includegraphics[width=0.333\textwidth,align=c,clip,trim={0.75cm 10.0cm 6.5cm 10.0cm}]{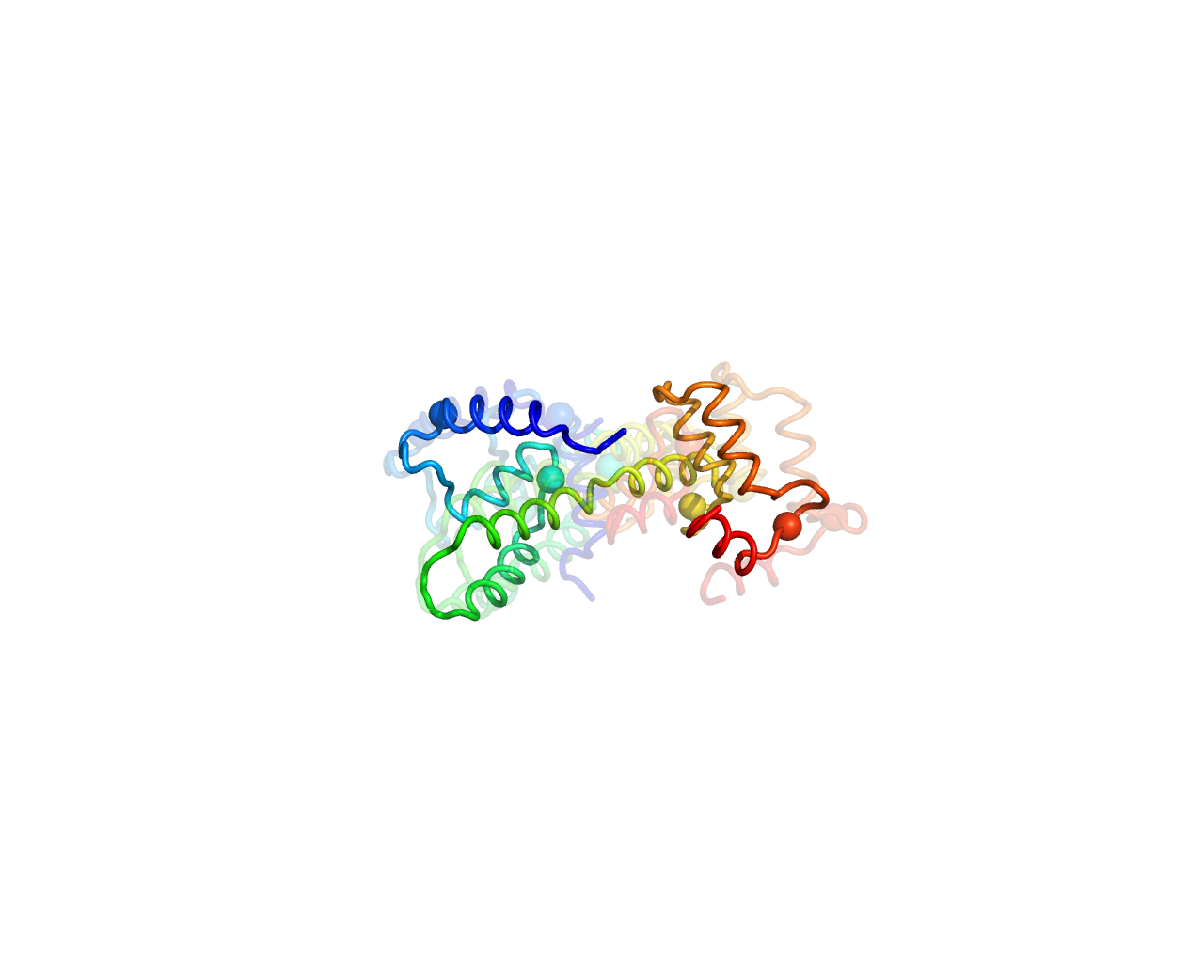}\\%
	\includegraphics[width=0.333\textwidth,align=c,clip,trim={4.0cm 10.0cm 6.5cm 10.0cm}]{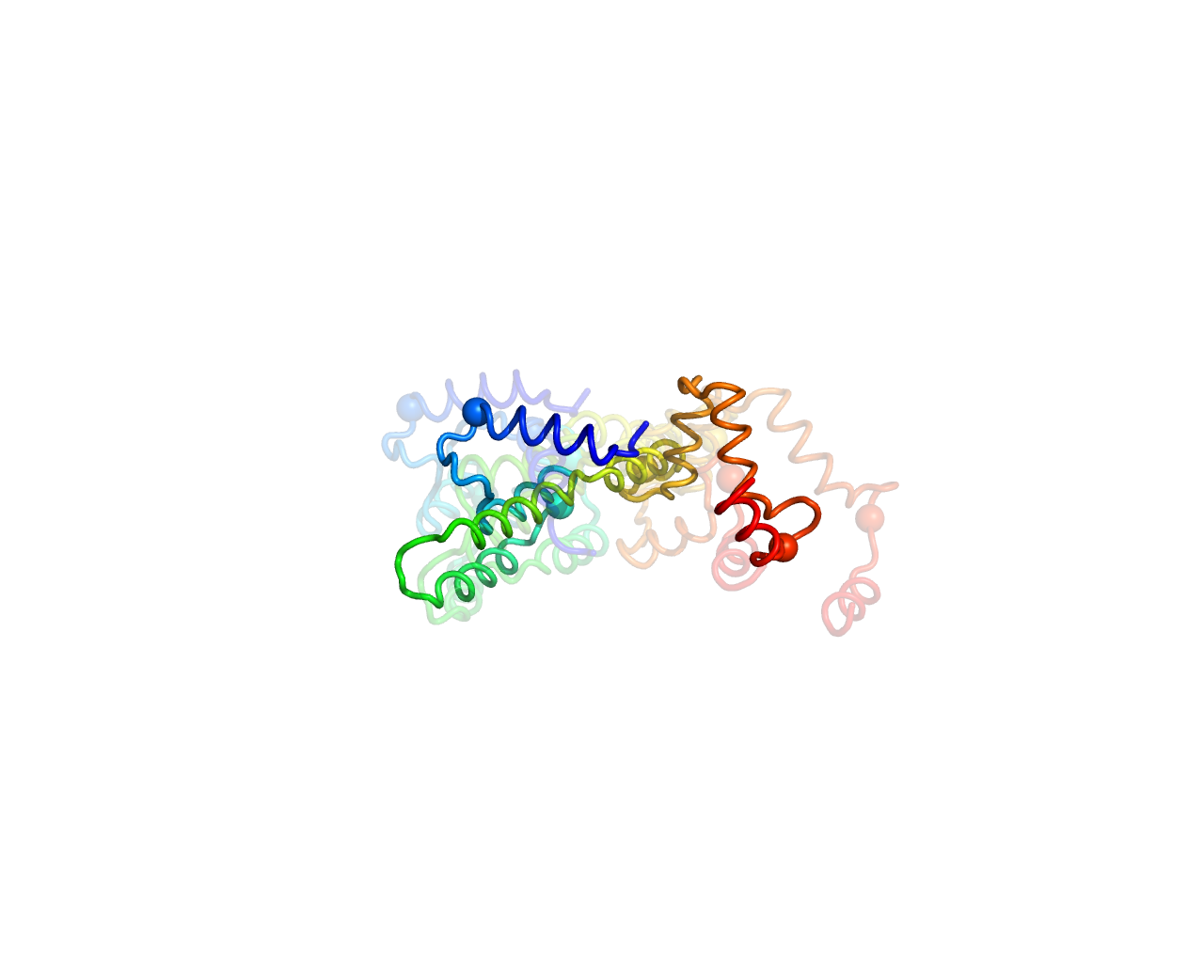}\includegraphics[width=0.333\textwidth,align=c,clip,trim={0.75cm 10.0cm 6.5cm 10.0cm}]{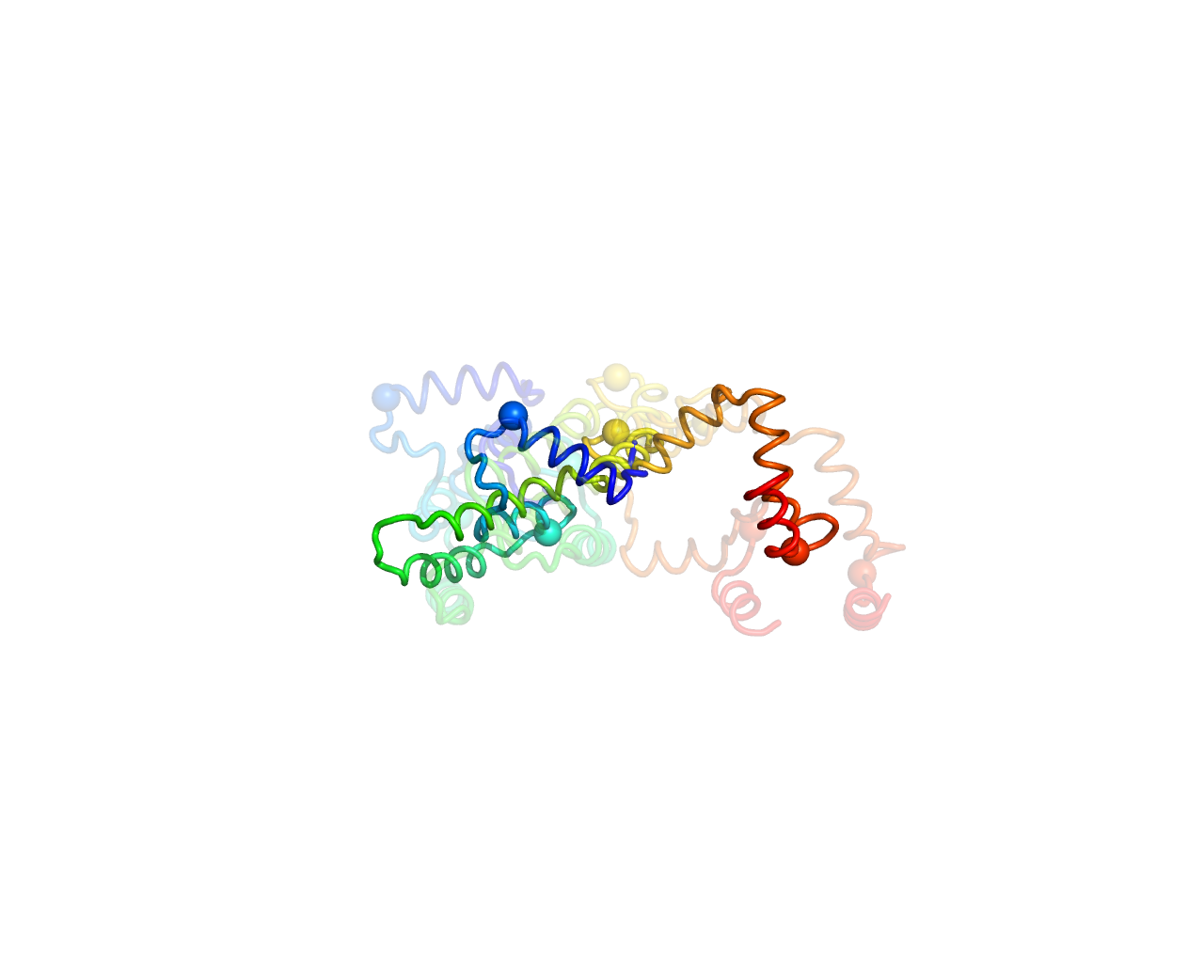}\includegraphics[width=0.333\textwidth,align=c,clip,trim={0.75cm 10.0cm 6.5cm 10.0cm}]{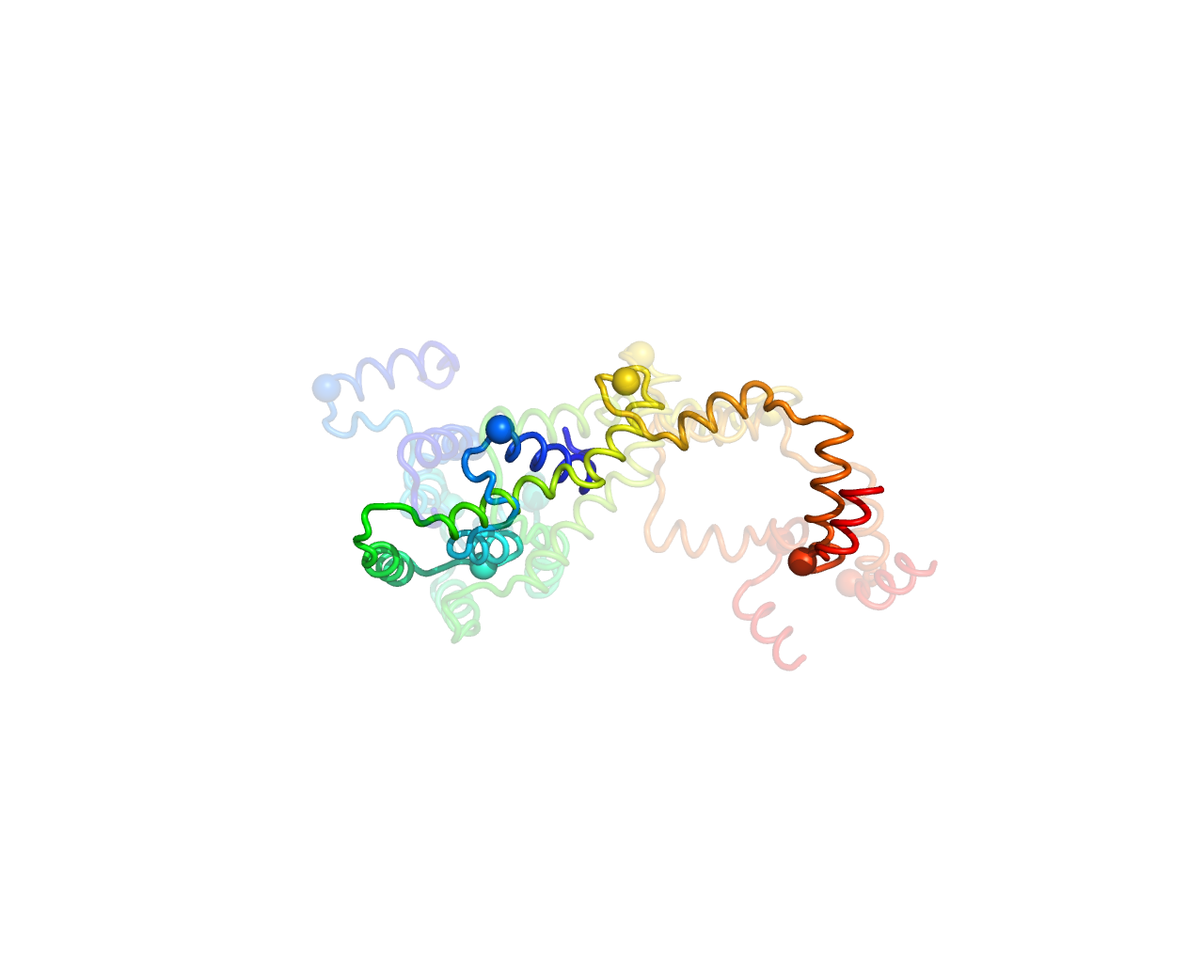}\\%
	\includegraphics[width=0.333\textwidth,align=c,clip,trim={4.0cm 10.0cm 6.5cm 10.0cm}]{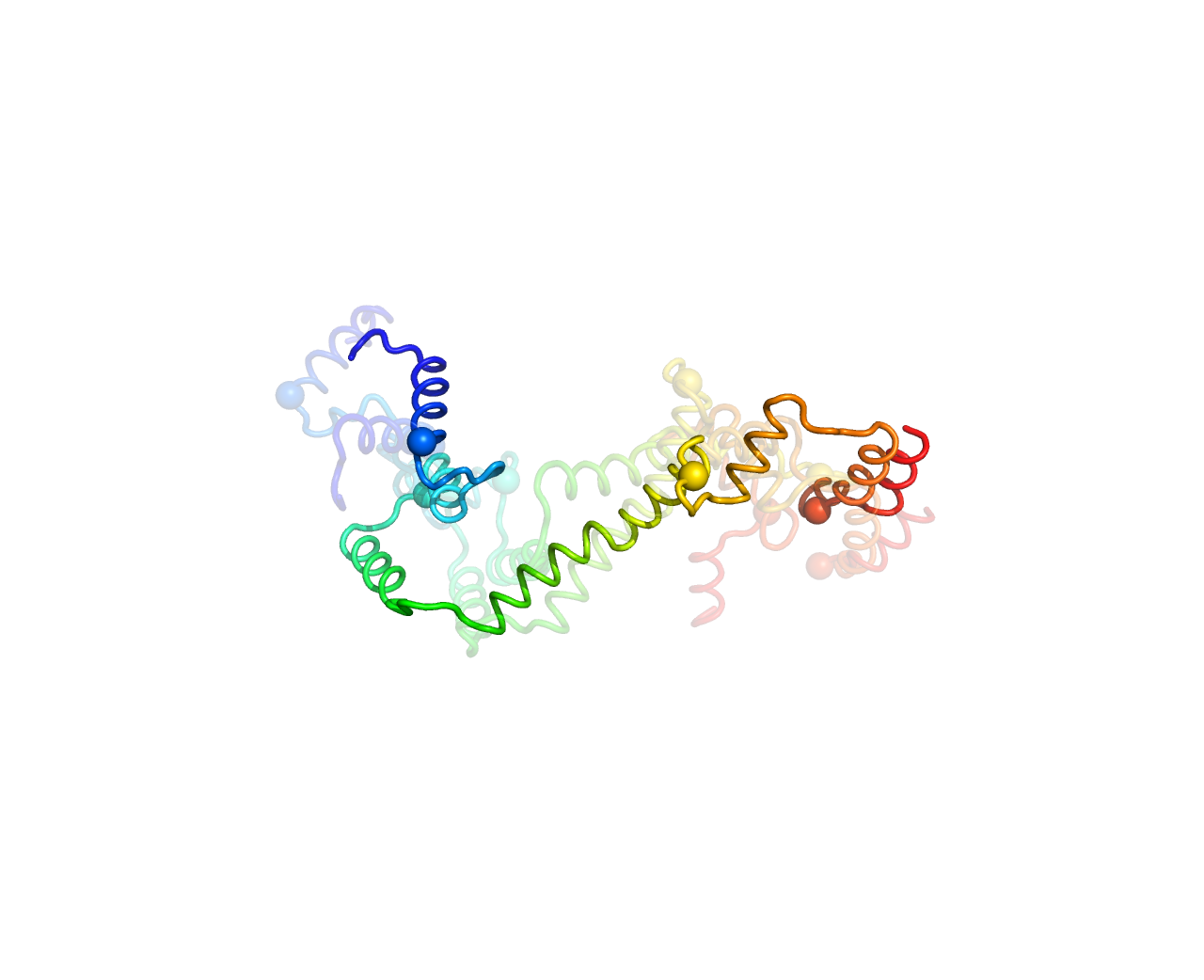}\includegraphics[width=0.333\textwidth,align=c,clip,trim={0.75cm 10.0cm 6.5cm 10.0cm}]{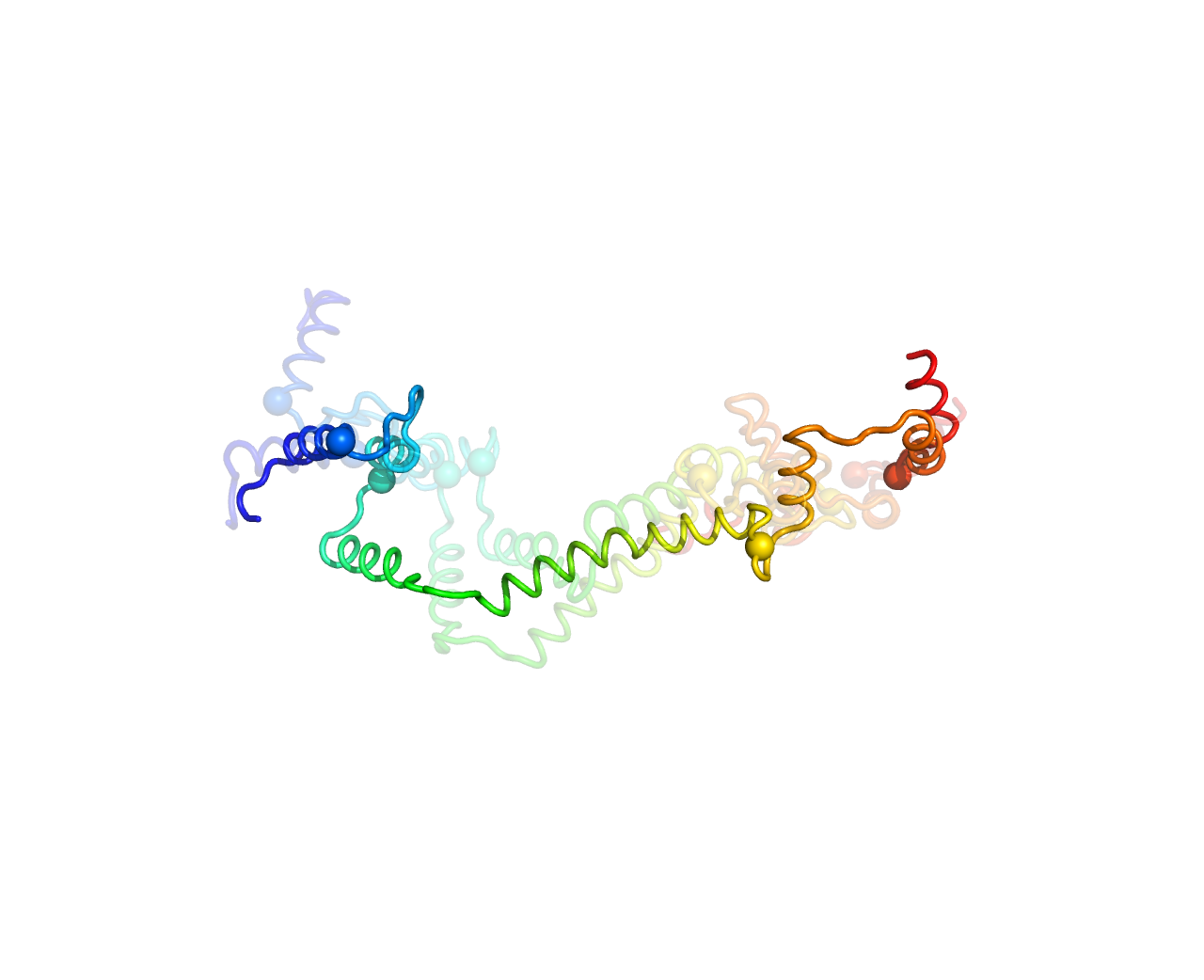}\includegraphics[width=0.333\textwidth,align=c,clip,trim={0.75cm 10.0cm 6.5cm 10.0cm}]{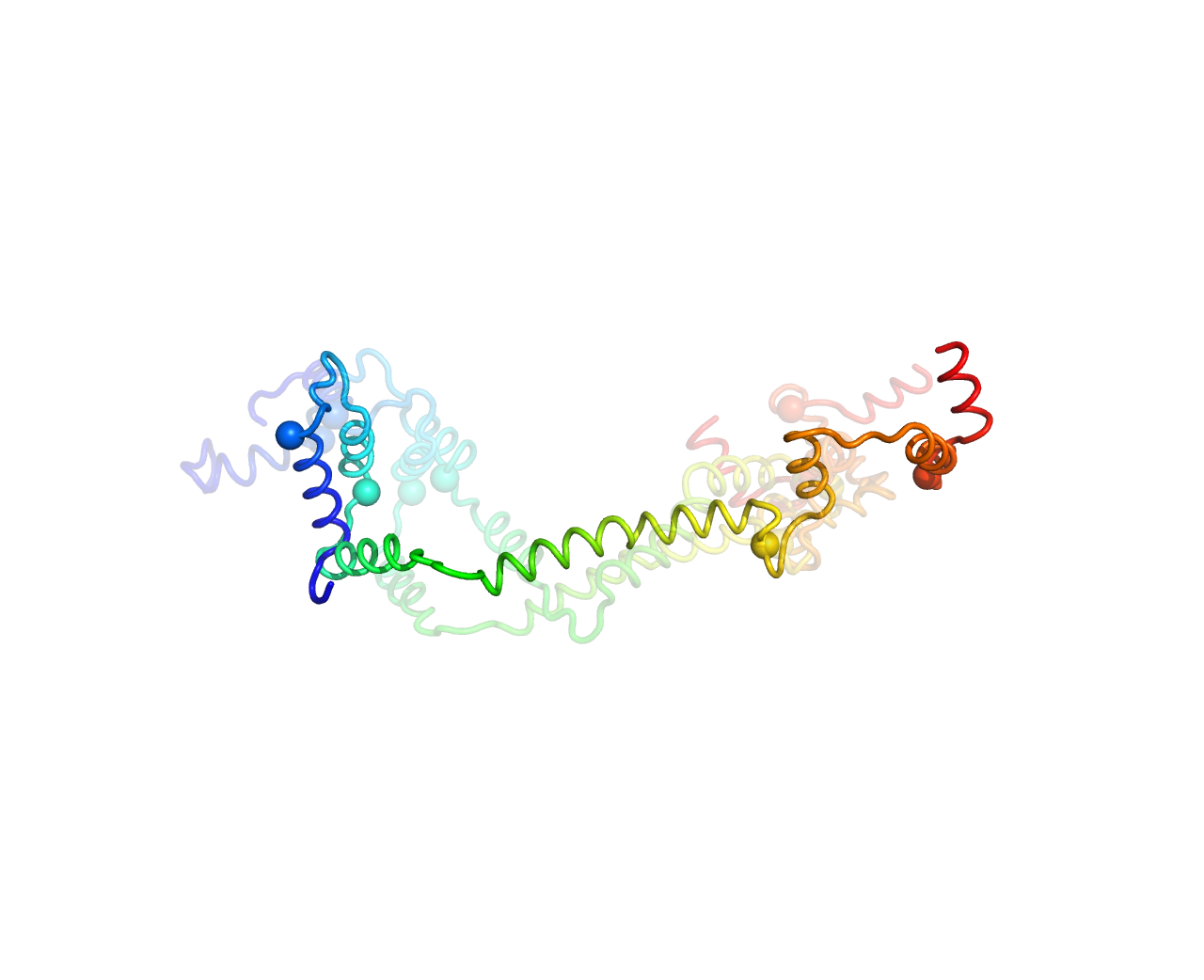}\\%
	\includegraphics[width=0.333\textwidth,align=c,clip,trim={4.0cm 10.0cm 6.5cm 10.0cm}]{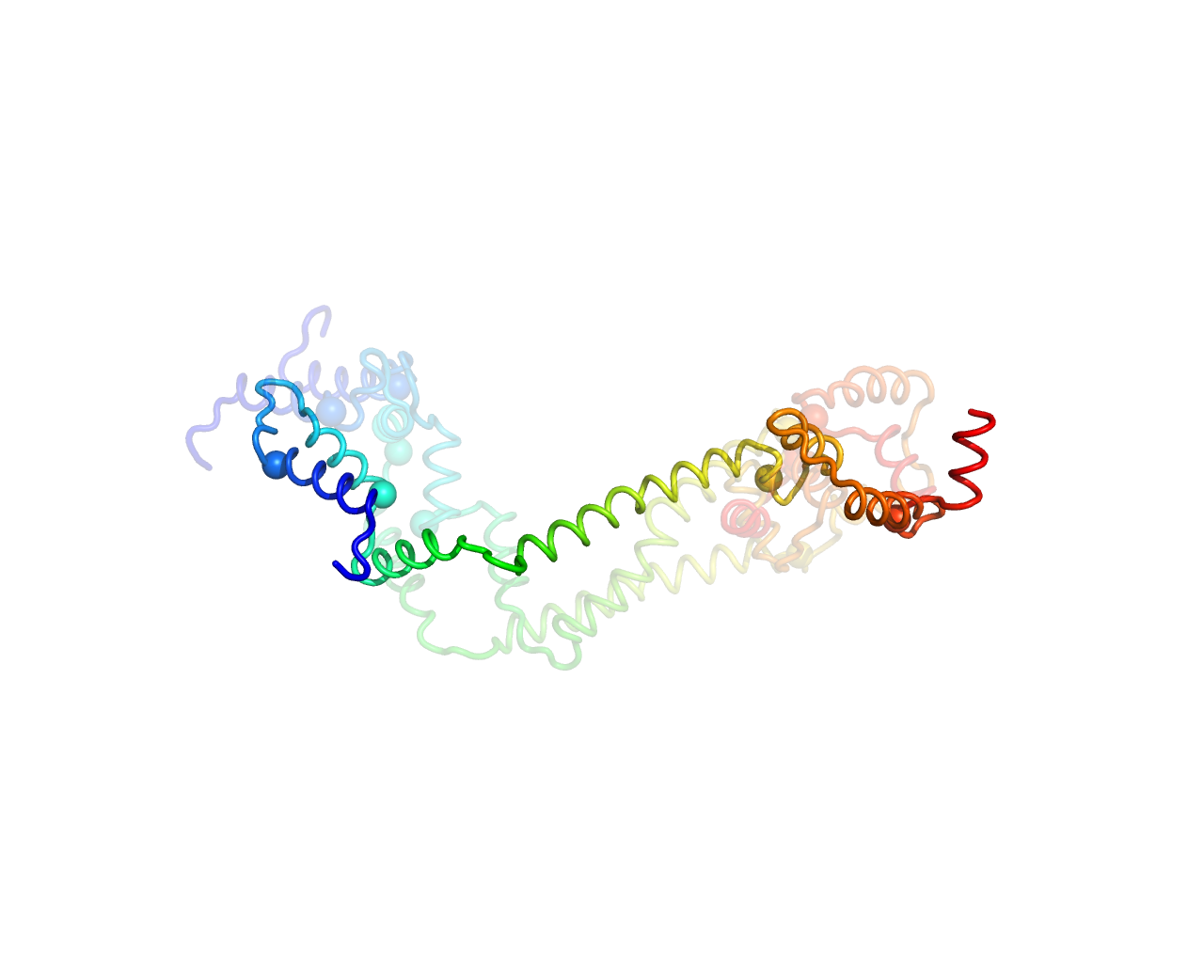}\includegraphics[width=0.333\textwidth,align=c,clip,trim={0.75cm 10.0cm 6.5cm 10.0cm}]{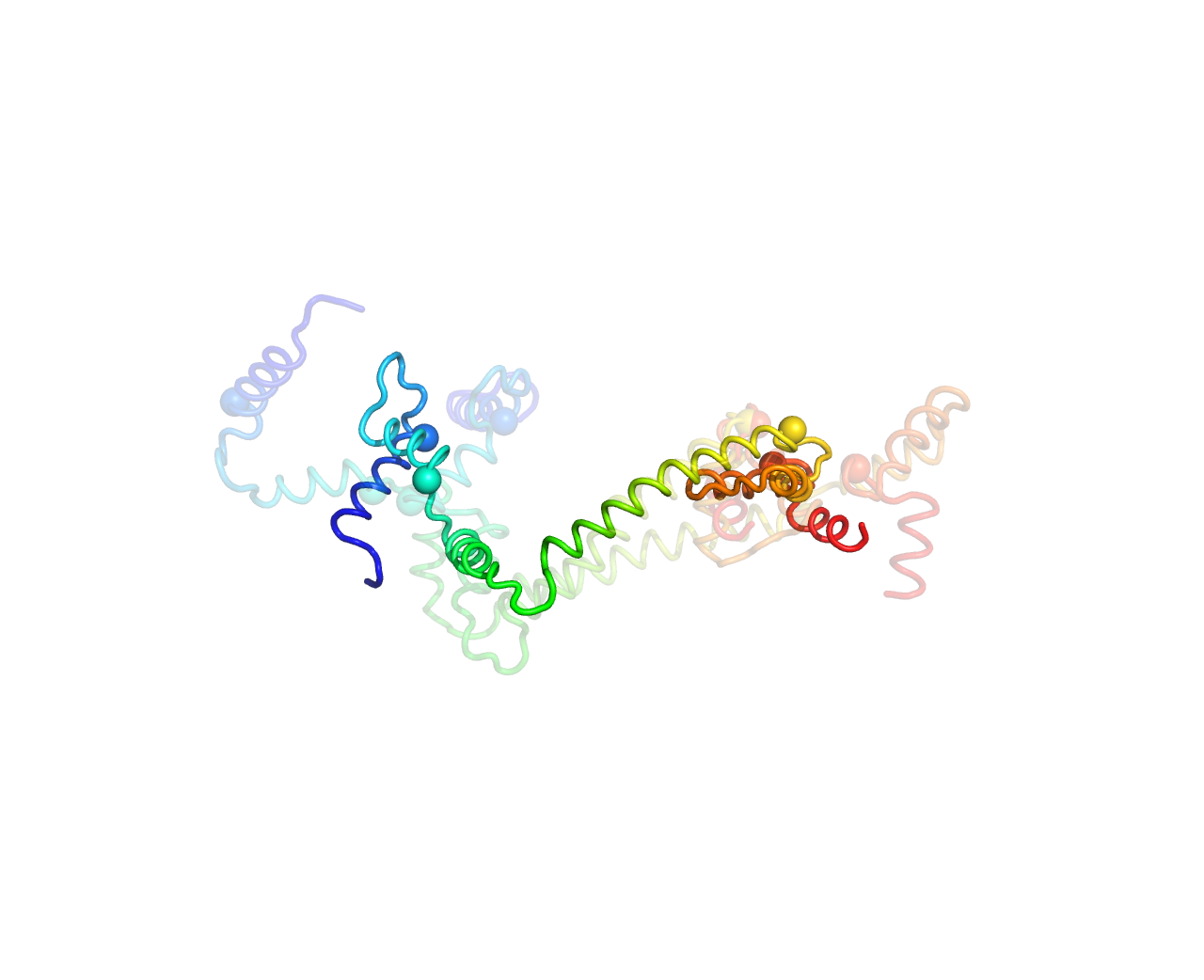}\includegraphics[width=0.333\textwidth,align=c,clip,trim={0.75cm 10.0cm 6.5cm 10.0cm}]{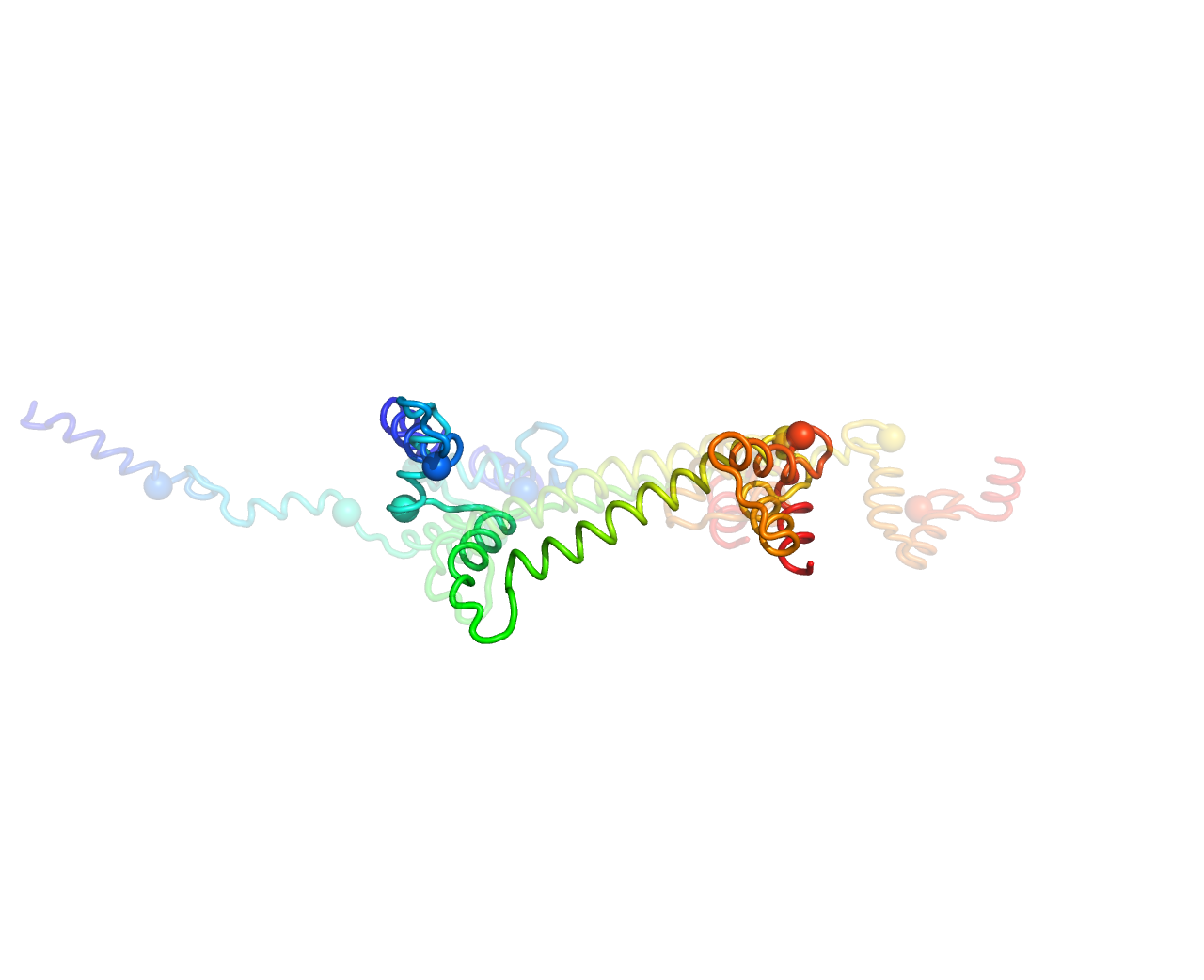}\\%
	\includegraphics[width=0.333\textwidth,align=c,clip,trim={4.0cm 10.0cm 6.5cm 10.0cm}]{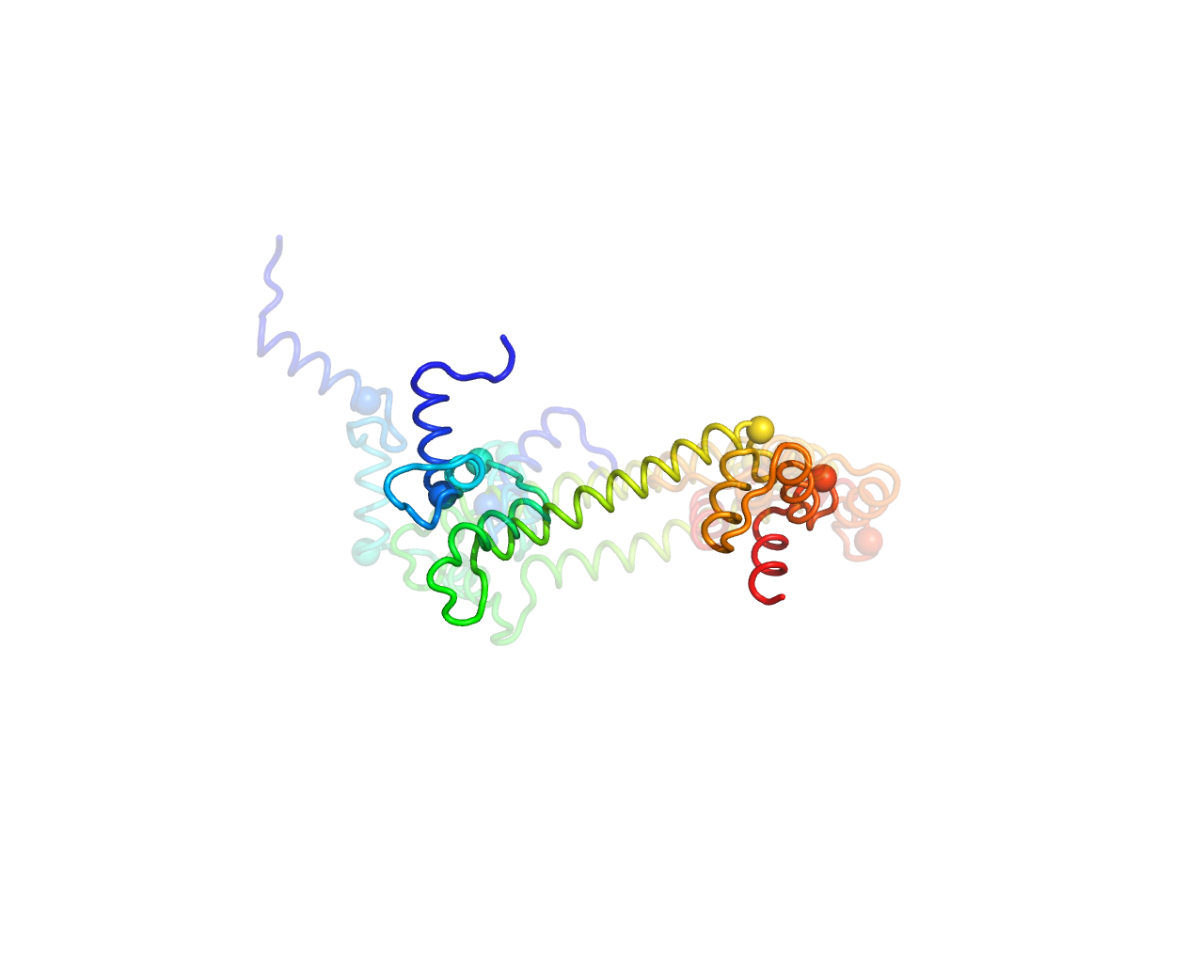}\includegraphics[width=0.333\textwidth,align=c,clip,trim={0.75cm 10.0cm 6.5cm 10.0cm}]{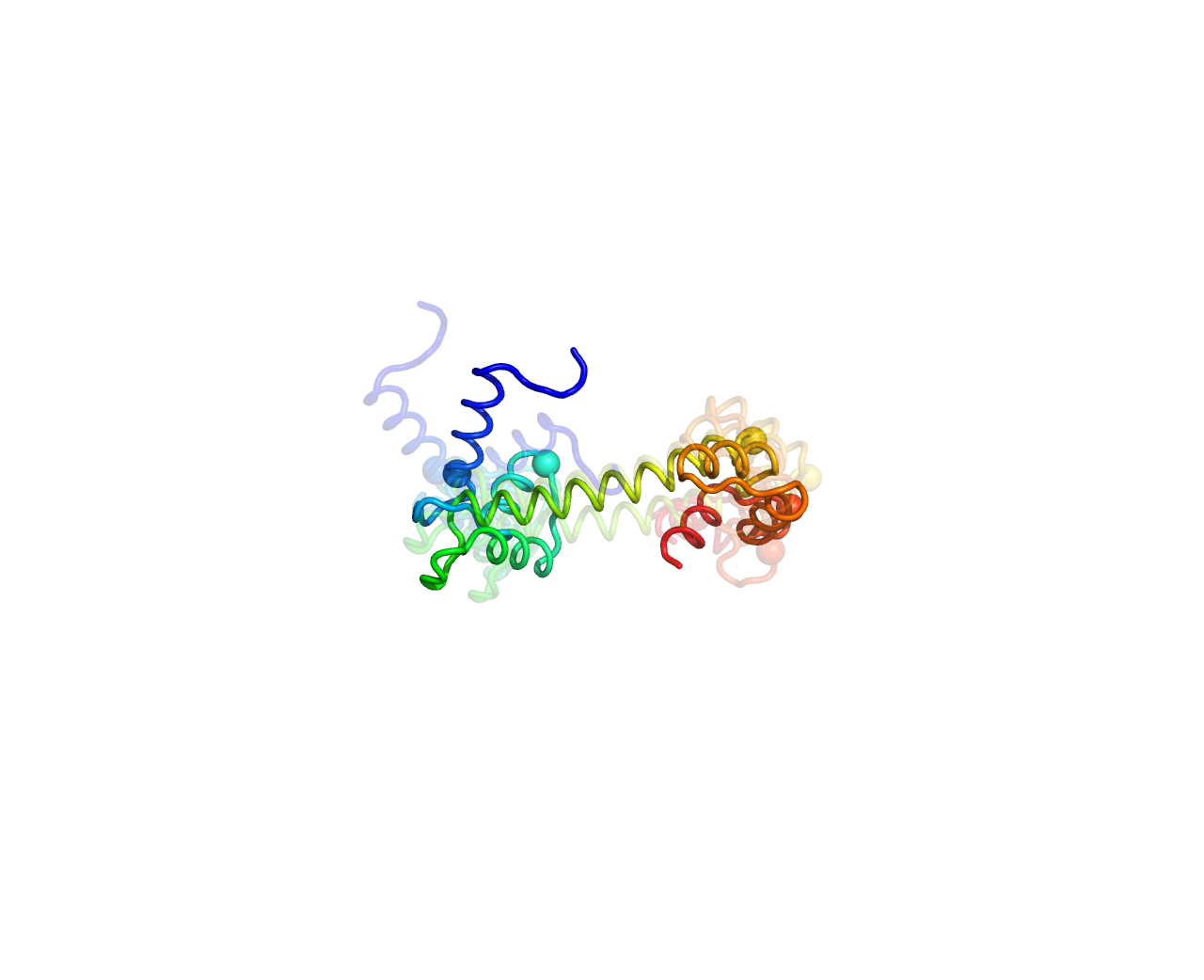}\includegraphics[width=0.333\textwidth,align=c,clip,trim={0.75cm 10.0cm 6.5cm 10.0cm}]{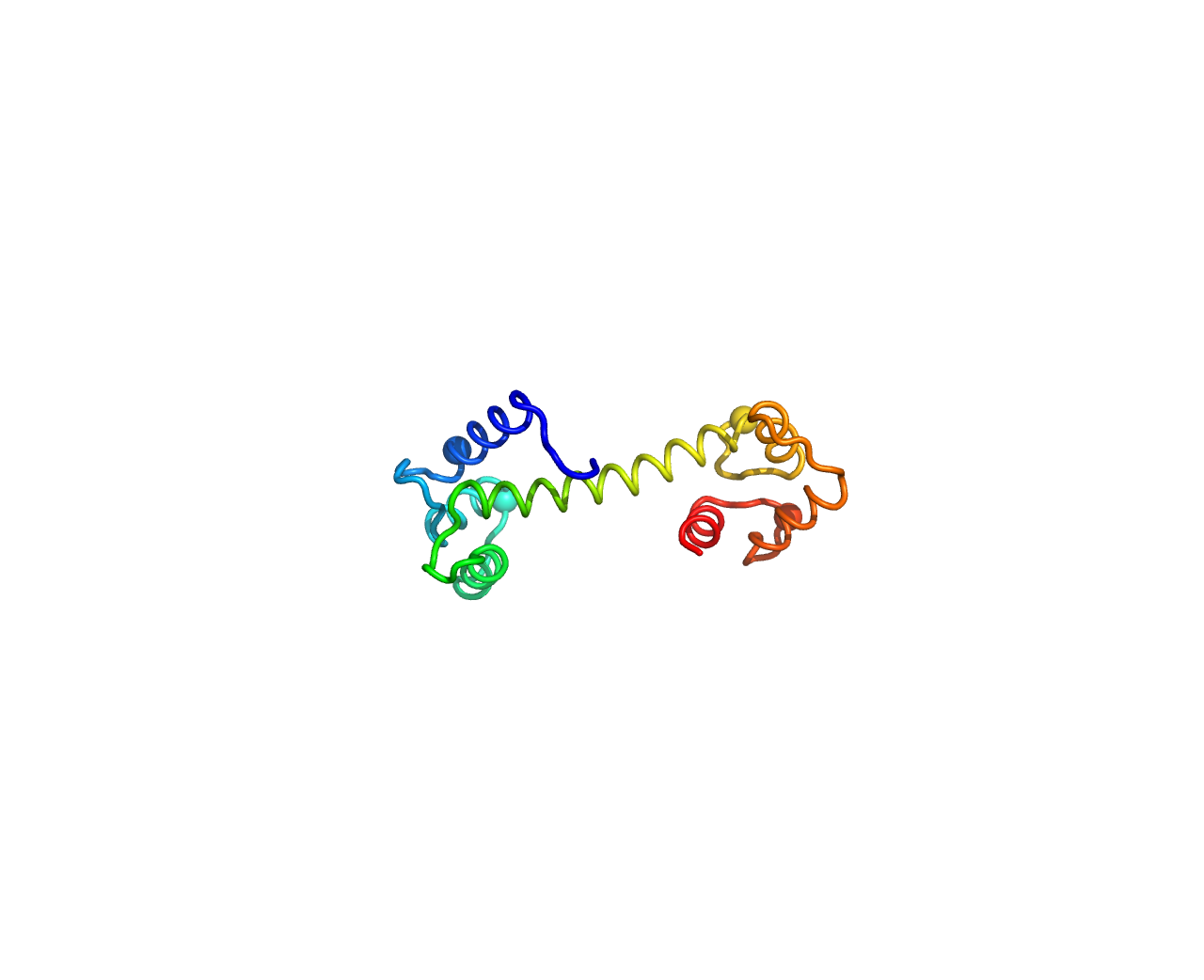}%
	\caption{\small From top to bottom and left to right, each figure represents a temporal snapshot of the three simulated protein bridges between the main chains of the proteins \texttt{1cfd} and \texttt{1rfj} according to the diffusion bridge described in \sect/\ref{sec:bridge}. The $15$ snapshots are at equidistant time points between the begin (\texttt{1cfd}, top left corner) and end (\texttt{1rfj}, bottom right corner) of the simulated bridges. The simulated trajectories shown with transparency correspond to $\alpha=0.25,1$, while the one shown in solid colours represents $\alpha=0.5$. The coloured spheres signal the positions of the amino acid sites $19$, $41$, $96$, and $135$. The corresponding three simulated dihedral bridges for these amino acids are shown in Figure \ref{fig:7}. The figure was prepared with PyMOL \citep{PyMOL}.}
	\label{fig:6}
\end{figure}

The present application gives a proof of concept on how toroidal diffusion bridges \eqref{eq:difftorus2} can be utilized to create protein bridges from estimated densities of pairs of dihedral angles. Further potential applications of these protein bridges include the simulation of the folding process of a protein $\mathsf{p}$ from its idealistic linear state with $\phi_i=\psi_i=0$, $i=1,\ldots,N$, to its folded state, or their use as stochastic interpolators in molecular dynamics simulations, allowing to refine to a higher temporal resolution a time series of protein structures $\mathsf{p}_{t_0},\ldots,\mathsf{p}_{t_n}$. \nowidow[4]

Several limitations exist in the proposed protein bridges that warrant further research. An important one is the independence between the diffusion processes associated with different amino acids and shared volatilities, which may induce intermediate unfolded shapes that are unrealistic. Dependence between diffusions could be introduced with a hidden Markov model that creates a density estimate on $\mathbb{T}^{2N}$ tailored to the amino acids in $\mathsf{p}_A$ and $\mathsf{p}_B$, as explored in \cite{Golden2017}. Additional challenges include avoiding potential atom clashes in intermediate backbones, and managing insertions and deletions of amino acids to handle proteins $\mathsf{p}_A$ and $\mathsf{p}_B$ with $N_A\neq N_B$. This last issue is needed to use these protein bridges to model protein structural evolution processes. In this regard, the presented bridges are time-reversible, aligning with the Pulley principle commonly considered in evolutionary trees \citep{Felsenstein1981}.

\section*{Acknowledgement}

The first author acknowledges support from grant PID2021-124051NB-I00, funded by MCIN/\-AEI/\-10.13039/\-50110001103 and by ``ERDF A way of making Europe''. His research was also supported by ``Convocatoria de la Universidad Carlos III de Madrid de Ayudas para la recualificación del sistema universitario español para 2021--2023'', funded by Spain's Ministerio de Ciencia, Innovación y Universidades. Part of this work was developed during the Oberwolfach workshop ``Statistics of Stochastic Differential Equations on Manifolds and Stratified Spaces''; both authors gratefully acknowledge the hospitality of the organizers and MFO. Part of the simulations were conducted at the Centro de Supercomputación de Galicia (CESGA). The authors appreciate data availability from Dr. Iago Sanmartín-Villar, as well as insightful discussions, regarding the application in \sect/\ref{sec:ants}. The authors also appreciate discussions with Prof. Thomas Hamelryck and Prof. Jotun Hein regarding the application in \sect/\ref{sec:prots}.


\appendix
%

\section{Proofs}

\begin{proof}{ of Proposition \ref{prp:Xt}}
	Set $g(x)\defin F^{-1}(x+F(\theta_0))$, $\theta_0\in\mathbb{T}^1$. Then $g'(x)=1/f(g(x))$ and $g''(x)=-f'(g(x))/f(g(x))^3$.
	Taking $X_t\defin g(\sigma W_t)$ and applying Itô's formula yields
	\begin{align}
		\rd X_t =\frac{\sigma^2}{2}g''(W_t)\rd t+ g'(W_t)\sigma\rd W_t%
		=-\frac{\sigma^2 f'(X_t)}{2f(X_t)^3}\rd t+ \frac{\sigma}{f(X_t)}\rd W_t.\label{eq:sde}
	\end{align}
	Therefore, $X_t$ is the solution of \eqref{eq:diff}, and hence $\Theta_t$ is by definition the solution of \eqref{eq:diffcirc}. The \pdf/ of $X_{t_2}\mid X_{t_1}=x_1$, denoted by $p_{t_2-t_1}^X$, follows by a direct transformation of variables:
	\begin{align*}
		p^X_{t_2-t_1}(x_2\mid x_1)%
		&=\phi_{(t_2-t_1)\sigma^2}(F(x_2) - F(x_1))f(x_2),\quad x_1,x_2\in\mathbb{R}.
	\end{align*}

	Consider a random variable $\Phi\defin m(X)$, $X\sim f_X$, where $m(x)\defin x\mod2\pi\in\mathbb{T}^1$, $x\in\mathbb{R}$, is a many-to-one function. The restriction $m_k\defin m\vert_{I_k}$, with $I_k=[2k\pi,2(k+1)\pi)$ and $k\in\mathbb{Z}$, has inverse $m_k^{-1}(\phi)=\phi+2k\pi$, $\phi\in\mathbb{T}^1$. Hence, the \pdf/ of $\Phi$ is given by
	\begin{align}
		f_\Phi(\phi)=\sum_{k\in\mathbb{Z}}f_X(m_k^{-1}(\phi))\frac{\mathrm{d}}{\mathrm{d}y}m_k^{-1}(\phi)=\sum_{k\in\mathbb{Z}}f_X(\phi+2k\pi),\quad \phi\in\mathbb{T}^1.\label{eq:wrapping}
	\end{align}
	After wrapping the diffusion, the \pdf/ of $\Theta_{t_2} \mid X_{t_1}=x_1$, denoted by $p_{t_2-t_1}^{\Theta,X}$, becomes
	\begin{align*}
		p_{t_2-t_1}^{\Theta,X}(\theta_2 \mid x_1)&=\sum_{k\in\mathbb{Z}} p_{t_2-t_1}^{X}(\theta_2+2k\pi\mid x_1)%
		=\sum_{k\in\mathbb{Z}} \phi_{(t_2-t_1)\sigma^2}(F(\theta_2)+k - F(x_1))f(\theta_2),\quad \theta_2\in\mathbb{T}^1.
	\end{align*}
	Now, for $\theta\in\mathbb{T}^1$, $p_{t_2-t_1}^{\Theta,X}(\theta_2 \mid \theta_1+2m\pi)=p_{t_2-t_1}^{\Theta,X}(\theta_2 \mid \theta_1)$,
	which implies that the distribution of $\Theta_{t_2} \mid \Theta_{t_1}=\theta_1$ equals that of $\Theta_{t_2} \mid X_{t_1}=\theta_1$. Thus, the \tpd/ of the circular diffusion \eqref{eq:diffcirc}, $p_{t_2-t_1}$,~is
	\begin{align*}
		p_{t_2-t_1}(\theta_2 \mid \theta_1)%
		&=2\pi f_{\mathrm{WN}}(2\pi F(\theta_2);2\pi F(\theta_1),4\pi^2(t_2-t_1)\sigma^2)f(\theta_2).
	\end{align*}
	Because $f_{\mathrm{WN}}(\theta;\mu,\sigma^2)$ converges to the uniform \pdf/ on $\mathbb{T}^1$ as $\sigma^2\to\infty$, it follows that
	\begin{align*}
		\lim_{t\to\infty}p_{t}(\theta \mid \theta_0)=f(\theta)2\pi\lim_{t\to\infty} f_{\mathrm{WN}}(2\pi F(\theta);2\pi F(\theta_0),4\pi^2 t\sigma^2)=f(\theta)
	\end{align*}
	for any $\theta,\theta_0\in\mathbb{T}^1$. The result concerning $f_{\mathrm{WN}}$ can be readily verified, e.g., by exploiting the Fourier form of the wrapped normal \pdf/: $f_{\mathrm{WN}}(\theta;\mu,\sigma^2)=\frac{1}{2 \pi}\left\{1+2 \sum_{k=1}^{\infty} e^{-k\sigma} \cos (k(\theta-\mu))\right\}$. This shows that $\{\Theta_t\}$ is ergodic with stationary \pdf/ $f$. Finally, it readily follows that $\{\Theta_t\}$ is time-reversible since $p_{t_2-t_1}(\theta_2\mid \theta_1)/p_{t_2-t_1}(\theta_1\mid \theta_2)=f(\theta_2)/f(\theta_1)$
	because $f_{\mathrm{WN}}(x;y,\sigma^2)=f_{\mathrm{WN}}(y;x,\sigma^2)$ for $x,y\in\mathbb{R}$.
\end{proof}

\begin{proof}{ of Proposition \ref{prp:Xt2}}
	We prove the first two statements, as the remaining ones follow analogously to the proof of Proposition \ref{prp:Xt}. We use the multidimensional Itô formula \citep[e.g.,][p. 48]{Oksendal2003}. Let $\bmV_t\defin\bmSigma^{1/2} \bmW_t$ and $\bmX_t\defin g(\bmV_t)$ for a $C^2$-function $g:\mathbb{R}^p\to\mathbb{R}^p$. Then, for $k=1,\ldots,p$,
	\begin{align*}
		\rd X_{k,t}
		&=\frac{1}{2} \sum_{i, j=1}^p \frac{\partial^{2} g_{k}(\bmV_t)}{\partial x_{i} \partial x_{j}} \Sigma_{i,j}\rd t + \sum_{i=1}^p \frac{\partial g_{k}(\bmV_t)}{\partial x_{i}} \rd V_{i,t}
		=\frac{1}{2} \mathrm{tr}[\bmSigma\mathrm{H} g_{k}(\bmV_t)]\rd t + \nabla g_{k}(\bmV_t) \bmSigma^{1/2} \rd \bmW_t%
	\end{align*}
	and therefore
	\begin{align*}
		\rd \bmX_t=\frac{1}{2} \begin{pmatrix}
			\mathrm{tr}[\bmSigma(\mathrm{H} g_{1})(g^{-1}(\bmX_t))] \\
			\vdots\\
			\mathrm{tr}[\bmSigma(\mathrm{H} g_{p})(g^{-1}(\bmX_t))]
		\end{pmatrix}\rd t + (\mathrm{D} g)(g^{-1}(\bmX_t))\bmSigma^{1/2}
		\rd \bmW_{t}.
	\end{align*}
	Setting $g(\bmx)\defin R^{-1}(\bmx+R(\bmtheta_0))$, $\bmtheta_0\in\mathbb{T}^p$, and $\bmTheta_t\defin g(\bmSigma^{1/2}\bmW_t)\mod 2\pi$ proves the first statement. The \pdf/ of $\bmX_{t_2}\mid \bmX_{t_1}=\bmx_1$, denoted by $p^{\bmX}_{t_2-t_1}$, follows by the transformation theorem as
	\begin{align*}
		p^{\bmX}_{t_2-t_1}(\bmx_2\mid \bmx_1)%
		&=\phi_{(t_2-t_1)\bmSigma}(R(\bmx_2) - R(\bmx_1))f(\bmx_2),\quad \bmx_1,\bmx_2\in\mathbb{R}^p,
	\end{align*}
	because $\left|\partial R(\bmx)/\partial \bmx\right|=f_1(x_1)f_2(x_2\mid x_1)\cdots f_p(x_p\mid x_1,\ldots,x_{p-1})=f(\bmx)$. Analogously to the proof of Proposition \ref{prp:Xt}, it follows that
	the \tpd/ of the the toroidal diffusion \eqref{eq:difftorus}, $p_{t_2-t_1}$, is
	\begin{align*}
		p_{t_2-t_1}(\bmtheta_2 \mid \bmtheta_1)=(2\pi)^p f_{\mathrm{WN}}(2\pi R(\bmtheta_2);2\pi R(\bmtheta_1),4\pi^2(t_2-t_1)\bmSigma)f(\bmtheta_2)%
	\end{align*}
	and that the process is time-reversible and ergodic with stationary \pdf/ $f$.
\end{proof}

\begin{proof}{ of Theorem \ref{prp:mle}}
	The first claim of the theorem follows from general results for estimating functions for ergodic stochastic processes in Theorem 3.2 in \cite{JacodSorensen2018}. Using basic calculus results and doing long and tedious calculations, it can be checked that under the conditions of the theorem, the estimating function $n^{-1} \partial_{\bmxi} \ell_n(\bmxi)$, i.e., a normalized version of the score function given by \eqref{eq:score}, satisfies Condition 3.1 in \cite{JacodSorensen2018}. The compactness of $\mathbb{T}^1$ is essential. In particular, it can be established that the classical interchange of integration and differentiation, which proves that $\mathrm{E}_{\bmxi_0}(\partial^2_{\bmxi} \log p_\Delta (\Theta_{2\Delta};\bmxi_0 \mid \Theta_\Delta)) = - \bmcalI^\Delta(\bmxi_0)$, is allowed.

	The asymptotic normality follows by standard arguments, see Theorem 2.11 in \cite{JacodSorensen2018}, from the central limit theorem for martingales, by which
	\begin{align}
		\label{eq:clt}
		n^{-1/2} \partial_{\bmxi} \ell_n(\bmxi_0) \stackrel{\cal D}{\longrightarrow} \mathcal{N}_q(\bzero,\bmcalI^\Delta(\bmxi_0)).
	\end{align}
	That the score function is a martingale follows by an interchange of differentiation and integration \cite[see, e.g.,][p.\ 11]{Sorensen2012}, which can be shown to be allowed under the conditions of the theorem. By the ergodic theorem, $\bmcalI^\Delta_n(\bmxi_0)$ converges in probability to $\bmcalI^\Delta(\bmxi_0)$. By arguments similar to those given in the proof of Theorem 3.2 in \cite{JacodSorensen2018}, it can be proved that the convergence of $\bmcalI^\Delta_n(\bmxi)$ is uniform for $\bmxi$ in a compact subset of $\Xi$ containing $\bmxi_0$. Thus $\bmcalI^\Delta_n(\hat\bmxi_n)$ converges to $\bmcalI^\Delta(\bmxi_0)$.

	Finally, \eqref{eq:Q} follows by standard arguments involving Taylor expansions; see, e.g., \cite{Silvey1970}. It is needed  that
	\smash{$-n^{-1}\ddot{\ell}_n(\hat\bmxi_n) \stackrel{\mathbb{P}_{\bmxi_0}}{\longrightarrow} \bmcalI^\Delta(\bmxi_0)$}, where $\ddot{\ell}_n(\bmxi)$ denotes the Hessian of $\ell_n$. This is true because $n^{-1}\ddot{\ell}_n(\bmxi)$ converges uniformly for $\bmxi$ in a compact set (see the proof of Theorem 3.2 in \cite{JacodSorensen2018}), and \smash{$-n^{-1}\ddot{\ell}_n(\bmxi_0) \stackrel{\mathbb{P}_{\bmxi_0}}{\longrightarrow} \bmcalI^\Delta(\bmxi_0)$}. In this way, it is found that $-2\log Q_n = \dot{\ell}_n(\bmxi_0)^\prime/\sqrt{n} \allowbreak\, \smash{\bmcalI^\Delta(\bmxi_0)^{-1/2} [\bmI_q - \bmR] [\bmI_q - \bmR ] \bmcalI^\Delta(\bmxi_0)^{-1/2} \dot{\ell}_n(\bmxi_0)/\sqrt{n} + o_P(1)}$,
	where $\smash{\bmR = \bmcalI^\Delta(\bmxi_0)^{1/2} \bmcalK(\bmxi_0) \bmcalI^\Delta(\bmxi_0)^{1/2}}$.\newline Here $\bmcalK(\bmxi_0)$ denotes the matrix with the upper left corner equal to $\smash{\bmcalI^\Delta(\bmxi_0)^{-1}_{11}}$ with $\smash{\bmcalI^\Delta(\bmxi_0)_{11}}$ denoting the upper left $q_1 \times q_1$-corner of the Fisher information matrix. Since $\bmR$ is idempotent of rank $q_1$, it follows that
	$\bmI_q - \bmR$ is the orthogonal projection on a subspace of dimension $q_2$. Thus the last result of the theorem follows from \eqref{eq:clt}.
\end{proof}

\begin{proof}{ of Theorem \ref{prp:anova}}
	As in the proof Theorem \ref{prp:mle} the result follows from general results in \cite{JacodSorensen2018} and Taylor expansions. First, we need to show that, as $n_iN_i \to \infty$, $i=1,\ldots, k$,
	\begin{align}
		\label{eq:clt2}
		\bmD_{\bmn,\bmN}^{-1/2} \, \partial_{\bmxi} \ell_{\bmn,\bmN}(\bmxi_0) \stackrel{\cal D}{\longrightarrow} \mathcal{N}_{kq}(\bzero,\bmcalJ(\bmxi_0)),
	\end{align}
	and \smash{$\bmD_{\bmn,\bmN}^{-1}\ddot{\ell}_{\bmn,\bmN}(\bmxi) = \bmD_{\bmn,\bmN}^{-1/2}\ddot{\ell}_{\bmn,\bmN}(\bmxi)\bmD_{\bmn,\bmN}^{-1/2}$} converges uniformly for $\bmxi$ in a compact set with limit $\bmcalJ(\bmxi_0)$ for the true parameter value $\bmxi_0$. Here $\bmD_{\bmn,\bmN} = \mbox{diag} (n_1N_1,\ldots,n_1N_1, \ldots , n_kN_k,\ldots,n_kN_k)$ with each $n_jN_j$ repeated $q$ times, while the Fisher information $\bmcalJ(\bmxi)$ is the block-diagonal matrix \smash{$\bmcalJ(\bmxi) = \mbox{diag}(\bmcalI^{\Delta_1}(\bmxi), \ldots ,\bmcalI^{\Delta_k}(\bmxi))$}.

	The assumption $n_iN_i \to\infty$ implies that either $n_i \to\infty$ or $N_i \to\infty$. For a group where $n_i \to\infty$, we can consider $N_i$ fixed and use that in the proof of Theorem \ref{prp:mle} we have shown that the score function for each observed trajectory is a martingale and checked Condition 3.1 in \cite{JacodSorensen2018}. Therefore we have previously established the limit results for the individual trajectories, and by independence the limit results follow for the group of trajectories.

	For a group where $N_i \to\infty$, we can consider $n_i$ fixed, and for this longitudinal data setting, we can use results in \sect/4 in \cite{JacodSorensen2018}. In our setting Condition 4.1 of that paper is essentially identical to Condition 3.1, which we checked in the proof of Theorem \ref{prp:mle}. Therefore, the limit results follows for such a group by the law of large numbers for i.i.d. variables, see the proof of Theorem 4.1 of \cite{JacodSorensen2018}. Thus the limit results have been established for all groups, and hence it follows that they hold for the score function of the entire data set.

	Under the assumptions of the theorem, $\bmD_0 \defin \lim (n_1N_1)^{-1}\bmD_{\bmn,\bmN}$ exists, and we can define \smash{$\tilde \bmM = \bmD_0^{1/2}\bmM$ and $\bmcalK(\bmxi_0) = \tilde \bmM (\tilde \bmM' \bmcalJ(\bmxi_0)\tilde \bmM)^{-1} \tilde \bmM'$}. By Taylor expansions and the second limit result above it is found that $\smash{\bmD^{1/2}_{\bmn,\bmN}(\hat \bmxi - \bmxi_0) = \bmcalJ(\bmxi_0) \bmD_{\bmn,\bmN}^{-1/2} \, \partial_{\bmxi} \ell_{\bmn,\bmN}(\bmxi_0) + o_P(1)}$ and $\smash{\bmD^{1/2}_{\bmn,\bmN}(\hat \bmxi_{\bmM} - \bmxi_0)} = \smash{\bmcalK(\bmxi_0) \bmD_{\bmn,\bmN}^{-1/2} \, \partial_{\bmxi} \ell_{\bmn,\bmN}(\bmxi_0)+ o_P(1)}$, where $\hat \bmxi_{\bmM}$ denotes the maximum likelihood estimator under the hypothesis $\bmxi \in \bmM A$. By further expansions we find that $\smash{-2\log Q_{\bmn,\bmN} = (D_{\bmn,\bmN}^{-1/2}\dot{\ell}_{\bmn,\bmN}(\bmxi_0))^\prime \, \bmcalJ(\bmxi_0)^{-1/2}} \allowbreak \smash{[\bmI_{kq} - \bmR] [\bmI_{kq} - \bmR ] \bmcalJ(\bmxi_0)^{-1/2} D_{\bmn,\bmN}^{-1/2} \dot{\ell}_{\bmn,\bmN}(\bmxi_0)/\sqrt{n} + o_P(1)}$ with \smash{$\bmR= \bmcalJ (\bmxi_0)^{1/2} \bmcalK (\bmxi_0) \bmcalJ(\bmxi_0)^{1/2}$}. Since $\bmR$ is idempotent of rank $\tilde q$, it follows that $\bmI_{kq} - \bmR$ is the orthogonal projection on a subspace of dimension $kq - \tilde{q}$. Thus \eqref{eq:Qlinear} follows from \eqref{eq:clt2}.
\end{proof}

\begin{proof}{ of Theorem \ref{prp:bridge}}
	By Proposition \ref{prp:Xt2}, %
	$\bmTheta_t=h(\bmW_t)$, where  $h(\bmx)\defin R^{-1}(\bmSigma^{1/2}\bmx+R(\bmtheta_0))\mod2\pi$, and $\{\bmW_t\}$ is a standard Wiener process (in particular $\bmW_0 =\bzero$). Note that $\bmTheta_0=\bmtheta_0$. The condition $\bmTheta_T=\bmtheta_T$ is equivalent to $\bmW_T=\bmSigma^{-1/2}[R(\bmtheta_T+2\bmk_T\pi)-R(\bmtheta_0)]=\bmSigma^{-1/2}[R(\bmtheta_T)-R(\bmtheta_0)+\bmk_T]$, with $\bmk_T$ in $\mathbb{Z}^p$. Let $\bmy_{\bmk}\defin\bmSigma^{-1/2}(R(\bmtheta_T)-R(\bmtheta_0)+\bmk)$, $\bmk\in\mathbb{Z}^p$, and let $A_i\subset\mathbb{T}^p$, $i=1,\ldots,n$, be arbitrary Borel sets. Then:
	\begin{align*}
		\mathbb{P}[\bmTheta_{t_1}&\in A_1, \ldots,\bmTheta_{t_n}\in A_n\mid \bmTheta_T=\bmtheta_T] \\
		&=\sum_{\bmk\in\mathbb{Z}^p} \mathbb{P}[\bmTheta_{t_1}\in A_1, \ldots,\bmTheta_{t_n}\in A_n\mid \bmK_T = \bmk,  \bmTheta_T=\bmtheta_T]\times\mathbb{P}[\bmK_T=\bmk\mid \bmTheta_T=\bmtheta_T]\\
		&= \sum_{\bmk\in\mathbb{Z}^p} \mathbb{P}[h(\bmW_{t_1})\in A_1, \ldots, h(\bmW_{t_n})\in A_n\mid \bmW_T=\bmy_{\bmk}] \times\mathbb{P}[\bmK_T=\bmk\mid \bmTheta_T=\bmtheta_T],
	\end{align*}
	where we have used that the events $\{\bmK_T = \bmk,  \bmTheta_T=\bmtheta_T\}$ and $\{\bmW_T=\bmy_{\bmk}\}$ are the same. It is not difficult to see that $\mathbb{P}[\bmK_T=\bmk\mid \bmTheta_T=\bmtheta_T]$ equals the right hand side of \eqref{eq:KT}. The theorem therefore follows because the distribution of $(\bmU_1,\ldots,\bmU_n)$ (constructed in Step \ref{bridge-step2}) equals the conditional distribution $(\bmW_{t_1},\ldots,\bmW_{t_n})\mid \bmW_T=\bmy_{\bmk}$. %
\end{proof}

\begin{proof}{ of Theorem \ref{prp:jumpbridge}}
	The theorem is proved in the same way as Theorem \ref{prp:bridge}. The main difference is that it must be proved that the distribution of $(U_1,\ldots,U_n)$ (constructed in Steps \ref{jumpbridge-step2} and \ref{jumpbridge-step3}) equals the conditional distribution $(V_{t_1},\ldots,V_{t_n})\mid V_T=y$, where $\{V_t\}$ is a Cauchy process with scale parameter~$\sigma$.

	The process $V_t = \tilde{W}_{S_t}$, where $\{\tilde{W}_t\}$ and $\{S_t\}$ are as explained above the theorem, is a (symmetric) Cauchy process with scale parameter $\sigma$. %
	Let $A_1, \ldots,A_{n+1}$ be independent $\mathcal{N}(0,2\Delta_i^{-2})$-distributed random variables. Then we can simulate the increments of $\{S_t\}$ by $S_{t_i}-S_{t_{i-1}} = A_i^{-2}$, $i=1,\ldots,n+1$, where $S_0=0$ by definition. With a slight abuse of notation, we denote the \pdf/ of $V_{t_1},\ldots,V_{t_{n+1}},A_1,\ldots,A_{n+1}$ by $f(v_1,\ldots,v_{n+1},a_1,\ldots,a_{n+1})$, and similarly for other densities. Then the \pdf/ of the bridge $V_{t_1},\ldots,V_{t_n} \mid V_{t_{n+1}}=v_{n+1}$ is
	\begin{align*}
		f(v_1,\ldots,v_n\mid v_{n+1}) &= \int_{(0,\infty)^{n+1}} f(v_1,\ldots,v_{n},a_1,\ldots,a_{n+1}\mid v_{n+1})\,\rd a_1\cdots \,\rd a_{n+1} \\
		&= \int_{(0,\infty)^{n+1}} f(v_1,\ldots,v_{n} \mid a_1,\ldots,a_{n+1}, v_{n+1}) f(a_1,\ldots,a_{n+1}\mid v_{n+1})\,\rd a_1\cdots \,\rd a_{n+1}.
	\end{align*}
	The first factor under the integral is the \pdf/ of a Brownian bridge with infinitesimal variance $2\sigma^2$ observed at the time points $\tau_i = A_1^{-2}+\cdots+A_i^{-2}$, $i=1,\ldots,n$ (simulated in Step \ref{jumpbridge-step3}). By Bayes' formula
	\begin{align*}
		f(a_1,\ldots,a_{n+1}\mid v_{n+1}) & =
		f(v_{n+1} \mid a_1,\ldots,a_{n+1}) f(a_1,\ldots,a_{n+1})/f(v_{n+1}) \\ & = \frac{\phi_{2\sigma^2(a_1^{-2}+\cdots+a_{n+1}^{-2})}(v_{n+1})}{f_{\mathrm{C}}(v_{n+1};0,(t_{n+1}\sigma)^2)} \prod_{i=1}^{n+1} \phi_{2\Delta_i^{-2}}(a_i).
	\end{align*}
	Thus $A_1,\ldots,A_{n+1}$ in Step \ref{jumpbridge-step2} is drawn from the conditional distribution given by the second factor under the second integral above with $v_{n+1}=y$. In conclusion, the vector $(U_1,\ldots,U_n)$ is a draw from the distribution of the Cauchy bridge $(V_{t_1},\ldots,V_{t_n})\mid V_T=y$.
\end{proof}
\end{document}